\documentclass{aa}

\usepackage{graphicx}
\usepackage{graphics}

\usepackage{color}

\begin{document}

\title{Pulsations, eruptions, and evolution of four yellow hypergiants.}

\titlerunning{Properties of yellow hypergiants}

\author{A.M. van Genderen\inst{1}
\and A. Lobel\inst{2}
\and H. Nieuwenhuijzen\inst{3}
\and G.W. Henry\inst{4}
\and C. de Jager\inst{5}
\and E. Blown\inst{6}
\and G. Di Scala\inst{7}
\and E.J. van Ballegoij\inst{8}
}
\institute {Leiden Observatory, Leiden University, Postbus 9513, 2300 RA Leiden, The Netherlands
 \and Royal Observatory of Belgium, Ringlaan 3, 1180 Brussels, Belgium
 \and SRON Laboratory for Space Research, Sorbonnelaan 2, 3584 CA Utrecht, The Netherlands
 \and Center of Excellence in Information Systems, Tennessee, State University, Nashville, TN, 37209, USA
 \and Royal Institute for Sea Research (Texel, The Netherlands), Molenstraat 22, 1791 DL Den Burg, Texel, The Netherlands
 \and 20 Cambridge Terrace, Masterton 5810, New Zealand
 \and Carner Hill Observatory, 9 Joshua Moore Drive, Horningsea Park, 2171, Sydney, Australia. Astronomical Association of Queensland, Brisbane, Australia
 \and KNWS Werkgroep Veranderlijke Sterren, De Rogge 6, 5384 XD Heesch, The Netherlands}

\date{Received..../Accepted....}

\abstract {}{We aim to explore the variable photometric and stellar properties of four yellow hypergiants (YHGs), HR\,8752, HR\,5171A, $\rho$\,Cas, and HD\,179821, and their pulsations of hundreds of days, and long-term variations (LTVs) of years. We also aim to explore light and colour curves for characteristics betraying evolutionary loops and eruptive episodes and to investigate trends of quasi-periods and the possible need for distance revisions.}{We tackled multi-colour and visual photometric data sets, looked for photometric indications betraying eruptions or enhanced mass-loss episodes, calculated stellar properties mainly using a previously published temperature calibration, and investigated the nature of LTVs and their influence on quasi-periods and stellar properties.} {Based on driven one-zone stellar oscillation models, the pulsations can be characterised as `weakly chaotic'.
The $BV$ photometry revealed a high-opacity layer in the atmospheres. When the temperature rises the mass loss increases as well, consequently, as the density of the high-opacity layer. As a result, the absorption in $B$ and $V$ grow. The absorption in $B$, presumably of the order of one to a few 0\fm1, is always higher than in $V$. This difference renders redder and variable $(B-V)$ colour indexes, but the absorption law is unknown. This property of YHGs is unpredictable and explains why spectroscopic temperatures (reddening independent) are always higher than photometric ones, but the difference decreases with the temperature. A new (weak) eruption of $\rho$\,Cas has been identified.
We propose shorter distances for $\rho$\,Cas and HR\,5171A than the accepted ones. Therefore, a correction to decrease the  blue luminescence of HR\,5171A by polycyclic aromatic hydrocarbon (PAH) molecules is necessary, and HR\,5171A would no longer be a member of the cluster Gum48d. HR\,5171A is only subject to one source of light variation, not by two as the literature suggests. Eruptive episodes (lasting one to two years), of YHGs prefer relatively cool circumstances when a red evolutionary loop (RL) has shifted the star to the red on the HR diagram. After the eruption, a blue loop evolution (BL) is triggered lasting one to a few decades. We claim that in addition to HR\,8752, also the other three YHGs have shown similar cycles over the last 70 years. This supports the suspicion that HD\,179821 might be a YHG (with a possible eruptive episode between 1925 and 1960). The range in temperature of these cyclic T$_{\rm eff}$ variations is 3000\,K--4000\,K. LTVs mainly consist of such BL and RL evolutions, which are responsible for a decrease and increase, respectively, of the quasi-periods. The reddening episode of HR\,5171A between 1960 and 1974 was most likely due to a red loop evolution, and the reddening after the 1975 eruption was likely due to a shell ejection, taking place simultaneously with a blue loop evolution. }{}


\maketitle

\section{Introduction}
In the present paper we explore large photometric data sets of four yellow hypergiants (YHGs) \object{HR\,8752\,=\,V509\,Cas}, \object{HR\,5171A\,=\,V766\,Cen,} and \object{$\rho$~Cas\,=\,HD\,224014} (the Big Three), and \object{HD\,179821\,=\,V1427 Aql}, the class of which is still undetermined (Arkhipova et al. 2009; Oudmaijer et al. 2009; Le Coroller et al. 2003; Ferguson \& Ueta 2010; Sahin et al. 2016). This latter could be a YHG or a red supergiant (RSG), but also a post-asymptotic giant branch(AGB) star, thus a much lower-mass object and has been incorporated into this study due to its spectroscopic and photometric similarities with both classes. The three types of objects, YHGs, RSGs and post-AGBs, are almost entirely convective.

Late-type supergiants showing the H$\alpha$ line with one or more broad emission components and exceptional broad absorption lines in their spectra are called hypergiants Ia$^{+}$. The approximate ranges for the M$_{\rm bol}$\,=\,-8.7\,--\,-9.6, for the luminosity log\,L/L$_{\rm \odot}$\,=\,5.3--5.7 and for the temperature T$_{\rm eff}$\,=\,4000\,K--6000\,K (excluding eruptive episodes). Typical for these stars are their very extended atmospheres, surface gravities around zero, high mass-loss rates, and strongly developed large-scale photospheric and atmospheric motion fields (de Jager 1980, 1998). The $\kappa$-mechanism of the two partially ionized He zones is likely responsible for the pulsations with a quasi-period of a few hundred days (e.g. Fadeyev 2011).
These stars are also in the state of gravitational contraction of their He core (Meynet et al. 1994; Stothers \& Chin 1996).

On a timescale of one to a few decades, a pulsation develops into an atmospheric eruptive episode, lasting about one to two years, reaching a high temperature (6500\,K--7500\,K; we note that this is not an evolutionary temperature rise). Lobel (2001) presented non-LTE calculations, showing that in this temperature range, cool supergiants become dynamically unstable due to the decrease of the first adiabatic index stability-integral $\langle$$\Gamma_{1}$$\rangle$ over a major fraction of the extended atmosphere as a result of partial, thermal and photo-ionization of hydrogen. At these high T$_{\rm eff}$ large fractions of partially ionized H gas
(T\,gas\,$\simeq$\,8000\,K) can synchronously recombine under pulsation decompression and drive the fast expansion of the entire atmosphere in an outburst event with strong global cooling. An enormous amount of energy is released which blows away part of the upper atmospheric layers.
Subsequently, a deep light minimum follows with T$\sim$\,4300\,K and the spectrum shows TiO absorption bands. The mass-loss rate as recorded during the  eruption of $\rho$\,Cas in the year 2000 amounted to $\sim$~3.10$^{-2}$\,M$_{\sun}$yr$^{-1}$ (Lobel et al. 2003). This amount is of the same order as that predicted by hydrodynamical models by Tuchman et al. (1978) and Stothers (1999).

Figure\,1 depicts a schematic HR diagram.
A cool portion of the HR diagram appeared to be almost void of stars (Humphreys 1978). Part of this void was theoretically studied and defined by de Jager (1998) and observationally checked by Nieuwenhuijzen \& de Jager (2000). Very relevant for this study was also the comparison of the evolutionary tracks of HR\,8752 and other YHGs with the models of Meynet et al. (1994) by Nieuwenhuijzen et al. (2012) in their Fig.\,1 and Sect.\,1.2. In this latter analysis,  IRC\,+10240 also appeared to be a YHG moving on an evolutionary track (Oudmaijer (1998).

This area was divided into two voids: the Yellow Void and the Yellow-Blue Void. These represent critical phases in the evolution of YHGs, bounded by the three semi-vertical curves. The curve at lowest T$_{\rm eff}$\,$\sim$~5200\,K is at photospheric values where H is close to beginning its ionization. The next curve at $\sim$\,7500\,K (dashed) is at photospheric values where H is close to being completely ionised and the curve at the highest T$_{\rm eff}$ between 8900\,K and 12500\,K marks the parameters where neutral He is close to being fully ionized.

The extreme photospheric values of the four YHGs studied in this paper are given in Table\,A:Data summary of the Appendix. It should be noted that with the exception of HR\,8752, no Hipparcos or Gaia Data Processing and Analysis Consortium (DPAC) parallaxes were used for the distances. Together with other experts in this field we advise against using these latter for very big stars and stars brighter than six, as then the reliability of parallaxes is very uncertain (van Leeuwen, priv. comm. 2019).

To prevent crowding in Fig.\,1, only one or two selected red or blue evolutionary loops (hereafter referred to as RL and BL, respectively) per YHG are schematically represented by lines (or dotted curves), and their observed evolutionary directions by an arrow.

For HR\,8752: a RL and a BL (based on Nieuwenhuijzen et al. 2012) are connected by the curved dotted line (during which the 1973 eruption happened) to indicate the evolutionary direction (not the true track).

\begin{figure}
 \resizebox{\hsize}{!}{\includegraphics{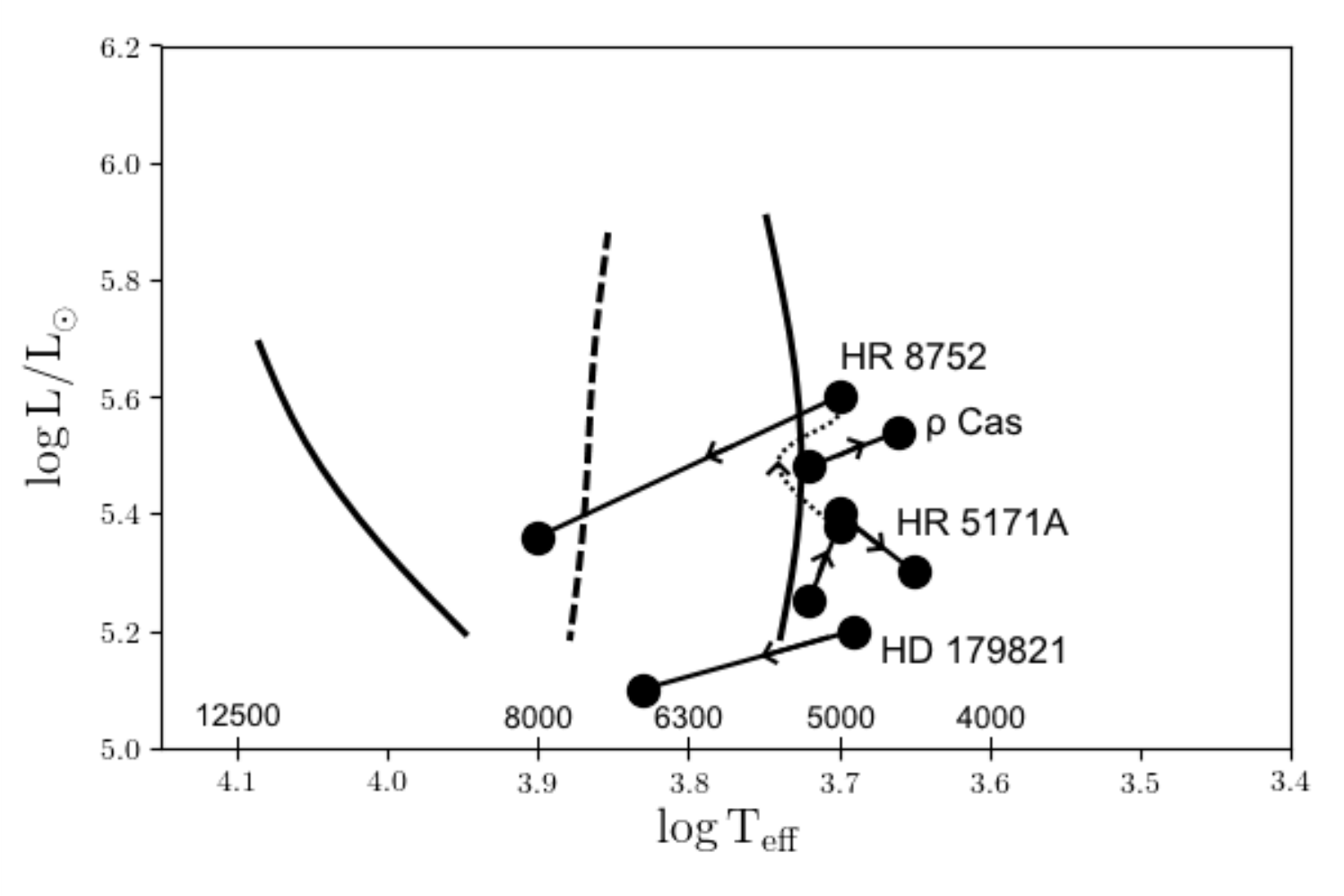}}
  \caption{A schematic HR diagram: on the right and left of the vertical dashed line the Yellow Void and the Yellow-Blue Void, respectively (data in Table\,A1:Data summary of the Appendix). The diagram shows the extreme physical and photometric values of one or two (HR\,8752) selected RL or BL evolutionary tracks of the program stars (arrows). Most of the extreme values are accidentally concentrated near the vertical curve on the right, the curve where H is close to beginning their ionization.}
  \label{Fig.1}
   \end{figure}

How long this zigzag behaviour lasts is unknown; it is meant to lose enough mass (mainly during the eruptive episodes) in order to evolve further to the blue to eventually become an S\,Dor variable (also referred to as a luminous blue variable (LBV)), or a Wolf-Rayet star (WR-star) (Oudmaijer et al. 2009; Oudmaijer \& de Wit 2013).

Critical comments by Sterken (2014, priv.comm.) on the supposition that HR\,5171A is a contact binary on account of a double-waved light curve constructed by Chesneau et al. (2014) contributed to our growing realization at the time (2014) that little was known about the photometric properties of the light curves of individual quasi-periodic pulsations and long-term light variations (LTVs) of YHGs.

Therefore, our focus was to probe the large photometric databases, comprising 50 to 120\,yr of observations of the four selected YHGs to the bottom, otherwise pivotal developments will be absent. The most important aims of this study are
to find out why spectroscopic temperatures are always higher than photometric ones; to decipher
whether or not in addition to HR\,8752 (Nieuwenhuijzen et al. 2012), the three other selected YHGs RL and BL evolutions on timescales of decades, interrupted by eruptive episodes;
to find out whether or not the observed LTVs represent these evolutionary tracks; and to find out whether the light variations of  HR\,5171A are due to pulsations alone, and why it reddened stubbornly from 1960 to 1981.
We also aim to find an explanation for the variable trends of quasi-periods.

Sections\,1 and 2 are devoted to the $\sim$\,1\,yr stellar instabilities and the stellar properties.
Section\,3 is devoted to the connection between LTVs and the evolutionary loops of YHGs on the HR diagram and the variable trends of quasi-periods. Section\,4 is devoted to a number of theoretical and speculative explanations for some observational highlights.
Section\,5 summarises the most important novel findings.

\section{Miscellaneous photometric and physical details and pulsation properties. }

\subsection{Databases and published extinctions.}

Based on decades up to more than a century of visual, mono-, and multi-colour photometry, it was obvious that the four selected objects show more or less the same pattern of 'semi-regularity' of pulsations, and appear coherent. The quasi-periods are a few hundred days in length and the pulsations are superimposed on the LTVs;
we refer to Appendix\,B:\,2.1. for references and descriptions.

Interstellar extinctions and reddening are required for our calculations of stellar properties: these are summarised in Table\,1.

\begin{table}
\caption{Adopted interstellar extinctions, reddening, and the extinction laws R for the four objects used in this paper. The reddening values, dependant on the extinction law R, in the Johnson $UBVRI$ were retrieved from the tables of Steenman \& Th\'e (1989).}
\label{table:1}
\centering
\begin{tabular}{c|c|c|c|c}
\hline\hline
 & HR\,8752 & $\rho$\,Cas & HR\,5171A & HD\,179821 \\
 \hline
 A$_{v}$ & 3.08$^{1}$ & 1.44$^{2}$ & 4.32$^{3}$ & 2.24$^{3}$ \\
 R &       4.4$^{1}$ & 3.2 $^{3}$ & 3.6$^{5}$ & 3.2$^{3}$ \\
 E$(B-V)$ & 0.70$^{1}$ & 0.45$^{4}$ & 1.2$^{3}$ & 0.7$^{6}$ \\
 E$(U-B)$ & 0.67     & 0.44              \\
E$(V-R)$ &  0.66         & 0.40     \\
E$(V-I)$  &  1.31           & 0.73     \\
E$(R-I)$  &  0.65          & 0.34     \\
\hline
\end{tabular}
\tablefoot{1\,=\,Nieuwenhuijzen et al. (2012); 2\,=\,Joshi \& Rautela (1978); 3\,=\,This paper, E$(B-V)$\,=\,0.4 for ISE and 0.8 for CSE; 4\,=\,Lobel et al. (2003); 5\,=\,van Genderen et al. (2015); 6\,=\,Arkhipova et al. (2009)}
\end{table}

\subsection{Pulsation properties. }

Figures\,2 and 3 show two typical time-series for the YHG cadence of pulsations. The first represents the very accurate $\rho$\,Cas photometry in $V$ and $B$ by Henry (1995, 1999) for the time interval 2003--2017 relative to the comparison star HD\,223173\,=\,HR\,9010 ($\triangle$ signifies variable minus comparison star). Henry's observations concern the T2 $VRI$ (1986--2001) and the T3 $BV$ (2003--2018) Automatic Telescope Projects (ATP). For the relative magnitudes we used for the comparison star: (Johnson system) $V$\,=\,5.513, $B$\,=\,7.160, $R$\,=\,4.220, $I$\,=\,3.390. See stub-tables Tables\,M.1. and M.2. in the Appendix for Henry's complete tables in electronic form available at the CDS.
The mean lines represent the LTVs. Between 2007 and 2011 (JD\,24\,54168--JD\,24\,55819) Klochkova et al. (2014) derived ten spectral temperatures (T(Sp)) of between 5777\,K and 6744\,K, thus showing a range of $\sim$\,1000\,K.

The two panels of Fig.\,3 show the cadence pattern of HR\,5171A (1989--1995 and 2009--2016, respectively), which is not significantly different from those of $\rho$\,Cas in Fig.\,2, of HR\,8752 (not depicted), and of HD\,179821 (e.g. Arkhipova et al. 2009).
Both panels of Fig.\,3 also show a significant declining visual brightness of the LTV.

In the upper panel of Fig.\,3 we matched professional photometry in $V$ based on several sources: van Genderen (1992, hereafter called Paper\,I); Chesneau et al. 2014) and visual observations by the American Association of Variable Star Observers (AAVSO) represented by the red curves which are based on the smoothed black dots. AAVSO data were shifted 0\fm4 upwards, the Hipparcos $H_p$ magnitudes shifted 0\fm1 downwards, and the $y$ ($uvby$) of the Str\"omgren system were shifted 0\fm1 upwards in order to match the $V$ of the Walraven $VBLUW$ system transformed into $V$ of the $UBV$ Johnson system (Paper I).

\begin{figure*}
 \sidecaption
  \includegraphics[width=12cm]{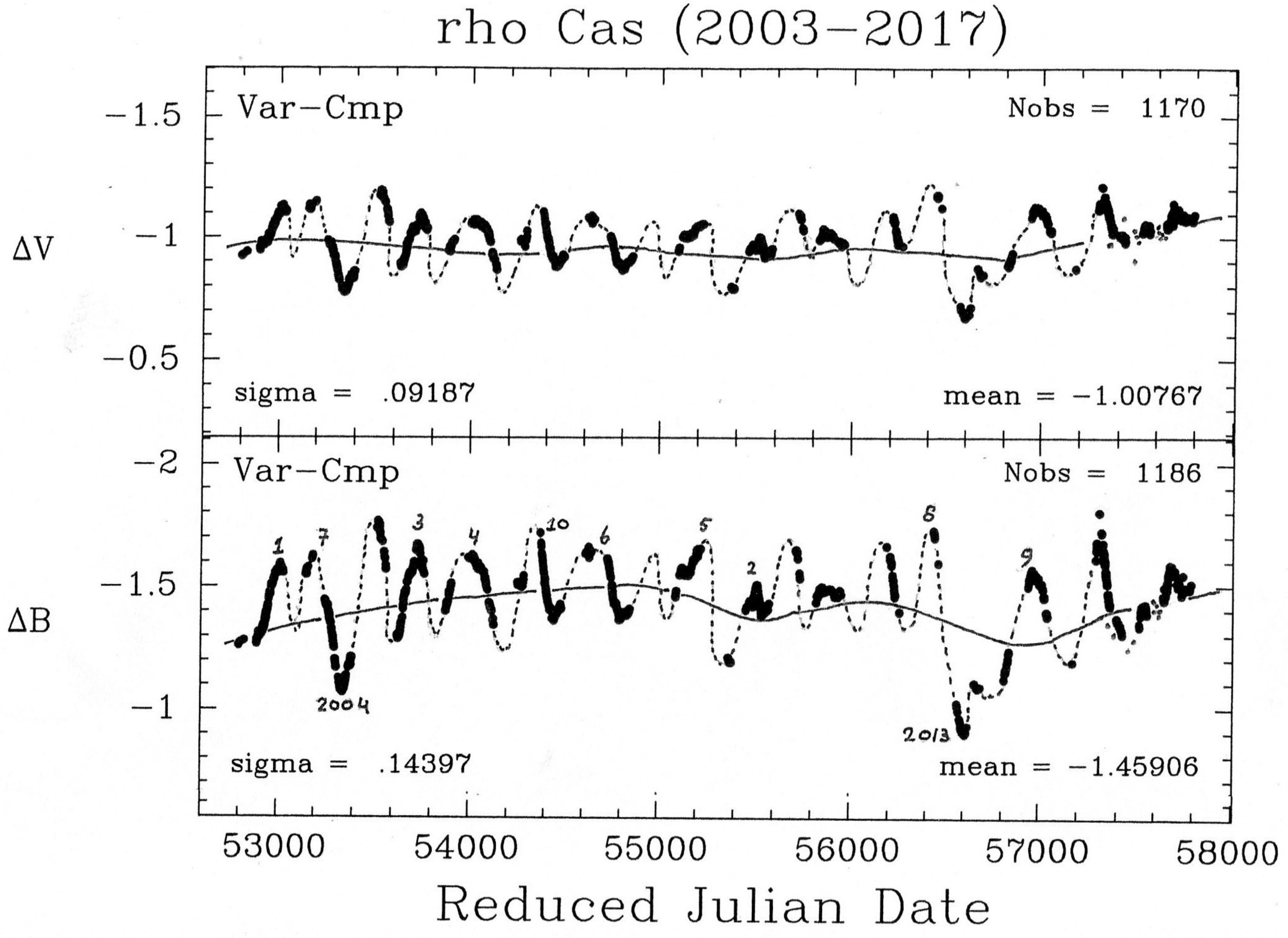}
   \caption{The 2003-2017 time-series in $B$ and $V$ ($UBV$) of $\rho$\,Cas relative to the comparison star by the T3 APT project (Henry 1995, 1999). Small dots on the right in $V$ and $B$ are observations made by one of the present authors (EJvB; see Appendix B: 2.1.). The mean curves sketched through the median magnitudes of the pulsations are the LTVs, in $V$ almost constant, in $B$ clearly present. Some numbers in the $B$ light curve represent pulsations quoted in the text. The sequence contains one new but weak eruption: 2013 (labelled no.8 in the $B$-panel and recognizable by its deep minimum). The pulsation in 2004, labelled no.\,7, has a relatively deep minimum, but is not an eruption episode. Spectral temperatures were derived by Klochkova et al. (2014) between JD\,24\,54168 and JD\,24\,55819.}
  \label{Fig2}
  \end{figure*}

No shift of the magnitude scale for the AAVSO data has been applied in the lower panel showing the difference with the six photo-electric $V$ magnitudes represented by plus signs (these $BV$ Johnson and $RI$ Cousins data are listed in Table\,I.1. of the Appendix). The systematic difference between the two observation techniques is common knowledge (Bailey 1978): the visual data are generally fainter by 0\fm2-0\fm4 than the photo-electric data (in the upper panel we applied a correction of 0\fm4).
The vertical line in the lower panel near JD\,24\,56755 represent an observation made with the aid of the Astronomical Multi-BEam combineR (AMBER) instrument of the VLTI instrument at the ESO by Wittkowski et al. (2017a), indicating that T$_{\rm eff}$\,=\,4290\,K\,$\pm$\,760\,K. We estimate that the brightness was $V$\,$\sim$\,6.4.

Our definition of the quasi-period, or duration $\triangle$\,D in days, is through timing measurements of light maxima. The amplitudes of the pulsations of the four objects vary between 0\fm2 and $\sim$\,0\fm5, and are variable from cycle to cycle.

Eruptions are striking, among other things, by the relatively large depth of the minimum ($\sim$\,1$^{\rm m}$) after the eruption maximum, called max1, of which the amplitude is sometimes slightly higher than for normal pulsation maxima. The duration of the entire eruption episode is about twice the duration of an ordinary pulsation. Because of such markers we identified one new eruption in Fig.\,2 ($B$ panel): in 2013 (the maximum named no.\,8). This eruption has been independently confirmed by spectroscopic observations (Aret et al. 2016). However, it was of a much weaker caliber than those of for example 1986 and 2000.
The pulsation in 2004, of which the maximum is labelled no.\,7 in Fig.\,2, showed more or less the same markers, but its light curve was even weaker than that of 2013, and it is definitely not an eruption (no spectroscopic indications).

\begin{figure*}
  \sidecaption
   \includegraphics[width=12cm]{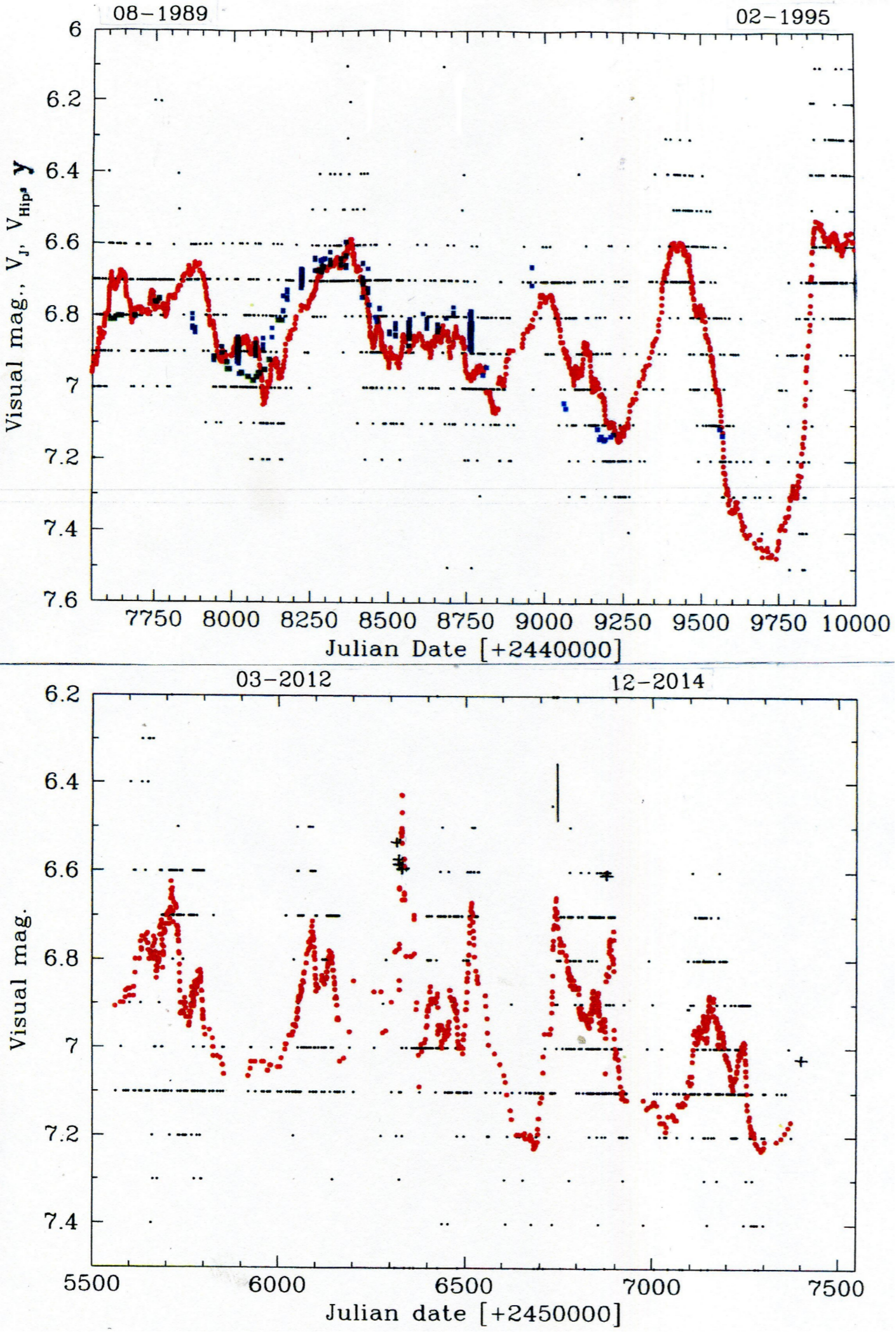}
 \caption{Upper panel:  1989--1995 pulsations of HR\,5171A in the visual (the averages of the AAVSO data shifted upward by 0\fm4 (black dots)
 from which smoothed averages were derived: the red dots). Other data points are added (e.g. Chesneau et al. 2014): the $V$ (Johnson system, transformed from the $V$ of the $VBLUW$ Walraven
    system, light green squares), the Hipparcos $Hp$ magnitudes shifted by 0\fm1 downwards (dark
   blue squares), and the $y$ ($uvby$) Str\"omgren system, shifted by 0\fm1 upwards (light blue, cyan squares). In this way all data fit the $V$ (Johnson system)
   magnitude scale satisfactorily. The deepest minimum on the right ($\sim$\,JD 244\,9750) is called the Jones-Williams minimum.
   The lower panel (we highlight the small scale difference between both JD axes) shows the 2010--2016 pulsations (based on AAVSO data). The last pulsation on the right with the minimum at $\sim$\,JD\,24\,57330 (early October 2015) is called the Blown-minimum. Six photo-electric
 $BV$ data (Johnson system) and $RI$ (Cousins system) observed by one of the present authors (GDS) are only for $V$ represented by a `+', and are listed in\,Table\,I.1. of the Appendix together with four more recent observations. No magnitude scale shift has been applied. The vertical line piece at JD\,24\,56755 (2014) marks the date of an AMBER/VLTI observation by Wittkovski et al. (2017a); see text. }
 \label{Fig3}
 \end{figure*}

There is some ambiguity about the two high peaks near JD\,24\,56500 in the lower panel of Fig.\,3, but because of the large scatter of the observations we considered them tentatively as a single peak. Errors for visual magnitudes of very red objects like HR\,5171A, are larger than those for less-red stars because of the Purkinje effect.
A quasi-period search in the literature is summarized in Appendix\,C:\,2.2.

After inspection of the large body of recorded photometric time-series of $\rho$\,Cas, HR\,5171A, HR\,8752, and HD\,179821, our conclusion is that YHGs are subject to a coherent sequence of pulsations, but that each pulsation is unique. These pulsations are quasi-periodic. Additionally, the quasi-periodicity is uninterrupted without any cessation of pulsations (including the eruptive pulsation episodes lasting twice as long). Only light curves of great precision reveal bumps, plateaus, and depressions lasting for weeks to months.

This incessant cyclic behaviour without any gaps in the sequence, which otherwise would signal one or more missing cycles (e.g. in case of one small-scale non-radial pulsation only happening accidentally on the invisible hemisphere), suggests that we are dealing with a more-or-less global instability pattern. In case of non-radial pulsations (Lobel et al. 1994; Fadeyev 2011), each hemisphere of the star should be well covered with a number of such oscillations moving up and down more or less in concert. The bumps and dips mentioned above could well be caused by this process.

\subsubsection{Semi-regular, weakly chaotic, or chaotic? }

The observations represented in Figs.\,2 and 3 (representative for the other two objects as well) show that YHGs are subject to coherent sequences of pulsations. One can easily identify minima and maxima, although showing a variable shape from cycle to cycle, without significant secondary dips and peaks. Periods and amplitudes of individual cycles vary by no more than a factor two or three (eruptive episodes excluded). Semi-regularity (semi\,=\,half) is an often-used but vague description for the variability of the different types of evolved stars. Below we summarise two of the most interesting studies dedicated to modeling the pulsation sequences of late-type stars and compare these latter with those of YHGs.

Following suggestions by Ashbrook et al. (1954) and Deeming (1970), Wisse (1979) used a continuous second-order auto-regression process to derive model light curves of red giants. The latter author distinguished two groups, one corresponding to strongly damped systems generating irregular (chaotic) model curves, and the other one consisting of slightly damped and undamped systems, generating light curves showing striking similarities with the real ones of Lb, SRb, and SRa giants; for example (i) the uniqueness of each cycle, (ii) a trace of a quasi-periodicity, and (iii) the presence of humps and shoulders. Wisse tentatively concluded that red giants are likely subject to random processes being dominant over one regular process.

Differently from their many predecessors, Icke et al. (1992) approached the study of such quasi-periodic behaviour with numerical studies of the dynamics of a driven one-zone stellar oscillation. These latter authors concentrated on the increase of the irregularity of the outer layers (the `mantle') of evolved AGB stars, which are low-mass stars. The more AGB stars evolve, the more mass they lose, and the less massive the mantle becomes, the more the oscillations become irregular.

We assume that the results of Wisse (1979) and Icke et al. (1992) are in general also valid for many other `semi-regular' convective variables such as YHGs, although these other variables are likely to be much more massive. Yellow Hypergiants have lost at least half of their initial mass, and are still highly eruptive on a timescale of decades, or less. Additionally, we assume that the mantle of YHGs will also have a lower mass compared to the total mass, similar to AGBs.

The calculations of the mantle motions by Icke et al. (which are assumed to be representative of the light curves), yield a number of global properties. A fine-looking series of oscillations may eventually, sometimes quite suddenly, switch to a different pattern.
We noted such sudden switches also in the mean quasi-periods of YHGs: after a number of oscillations, for example about half a dozen, the mean jumps to a smaller (or larger) one.

With respect to this type of intermittent behaviour, Icke et al.: state that if a stellar oscillator is found to be intermittent, an impractical amount of observing time is needed to establish the periodicity. This is exactly the case for YHGs (apart from the sudden switches of the mean quasi-periods mentioned above); each pulsation has a different duration. Additionally, we observe that also with respect to morphology, each pulsation is different as well. In short, each pulsation is unique, and predicting any property of the next pulsation is impossible.

Icke et al. also concluded that stars with a large interior radius relative to that of the mantle likely show a more chaotic behaviour. For YHGs the opposite is likely the case: the outer layers are extremely extended, thus reducing the chaotic character.

Noteworthy is that for the calculation of the model light curves of different kinds of evolved late-type stars, Wisse (1979) and Icke et al. (1992) did not take into account any potential long-term variation; Sect.\,4.

In conclusion, we characterise the instability of the four objects studied here as weakly chaotic, that is, not as chaotic as the individual oscillations that are clearly separated from each other. In other words, the cadence of oscillations appears to be coherent.

\subsection{HR\,5171A: a contact binary?}

One of the important purposes of Figs.\,2 and 3 was to emphasise that despite the weakly chaotic character, the quasi-periodic sequences of YHGs show a coherent behaviour. Therefore, it was easy to conclude that HR\,5171A does not show any trace of a second variable source as proposed by Chesneau et al. (2014). The light curve of these latter authors consists of a double-peaked brightness oscillation with two unequal amplitudes, and with a supposed binary period of about 1300 d. This is unlikely, as if this were the case Fig.\,3 would not show any coherent succession of light oscillations, but a quite chaotic distribution of the observations. The resulting brightness fluctuations would vary from very small to twice the present size (Fig.\,3), by two oscillations with slightly different amplitudes, but significantly different periods (the one with half the binary period: about 660 d, and the other one by pulsation, at the time about 500 d). With respect to the morphology of the observed light curves observed in Figs.\,2 and 3, it happens to be that pulsations of $\rho$\,Cas and HR\,5171A showed minima with alternating depths for some time. The selection of cycles as done by Chesneau et al. and depicted in their Fig.\,8 was obviously sensitive to this bias, resulting in the double wave with two unequal minima. However, there are mitigating circumstances: at the time very little was known about this treacherous photometric property (hopefully, the current paper fills this gap).

Another argument in favour of pulsations only is that the monitoring of the V$_{\rm rad}$ curve during two pulsations (no.\,16 and no. 17 in Fig.\,1 of Paper \.I) by Lobel et al. (2015) showed a regular phase lag of about 0.4 with respect to the light curve. This is also typical for $\rho$\,Cas (Lobel et al. 2003).

However, we do not deny the possible presence of some companion, albeit it perhaps only in the field of view of the Hipparcos satellite; see the photometric analysis by Eyer (1998) and footnote no.\,2 in van Genderen et al. (2015). Based on AMBER/VLTI observations, Chesneau et al. (2014) discovered a bright spot in front of the primary disk. The PIONIER/VLTI interferometric observations by Wittkovski et al. (2017b, c) obtained impressive near-infrared (NIR) $H$-band images. They claim that the contact companion, suggested by Chesneau et al. (2014) is tentatively supported by a few images of HR\,5171A showing no precise circular circumference. One image even showed a clear bubble in the SW direction, which according to these authors might be the companion. A tentative model for the orbit is proposed. Secondary features like bright patches are assigned to a convection cell.

It can also be concluded that if a low-mass contact companion to HR\,5171A were present, any geometrical light variation due to deformation is obviously too small to be clearly detected.
The diameters of those few images mentioned above of HR\,5171A look different from each other, obviously due to the pulsations (light curve at the time shown in Fig.\,15), and the circumference is far from perfectly circular. This is understandable considering the very low surface gravity.

Therefore, we also conclude that the well-covered and clearly plotted pulsation sequences like those in Figs.\,2 and 3, are essential to understand the characteristic instabilities of YHGs and possibly to correct questionable interpretations.

\subsection{Inconsistencies between brightness and colour, and between spectroscopic and photometric temperatures.}
The analyses presented in the following sections have not been performed previously.
We embarked on the (tedious) traditional method of plotting many tens of individual pulsations per object on graphic paper, comprising thousands of data points (for light and colour curves). The intention was to get an idea about the degree of their variable morphological properties, whether quasi-periods and light amplitudes are different during hot and cool stages, and more importantly to try to understand why spectroscopic and photometric temperatures, T(Sp) and T(Phot), always differ. Pulsations during which a T(Sp) was obtained were particularly important for our purposes.

Walker (1983), studying photometry and spectroscopy of HR\,8752, was probably the first to notice that a relationship between the photometric magnitudes (and thus the continuum radiation) and the spectrum of YHGs was not always straightforward. He was not the only one.
For example, Arkhipova et al. (2009) noted ambiguous relations between brightness and colours of the HD\,179821 pulsations during 2000--2009, and attributed them to increasing optical depth during high mass-loss episodes.

We also noticed that differences exist between the temperatures based on spectroscopy and photometry: large and small for high and low stellar temperatures, respectively. Such differences not only exist for $\rho$\,Cas, but also for the three other objects.

The purpose of the following sections is to enhance our understanding of the photometric properties and the probable cause of the fact that T(Sp) is usually higher than T(Phot).

\subsection{Correlation diagram $V$/$(B-V)$: similarities and differences between hot and cool YHGs.}
Studying the morphology of the series of detailed light and colour curves provides valuable information about the wavelength dependency of amplitudes, the variation of the duration of pulsations, the size of amplitudes, and the mean rate of variation of temperature and radius. Enhanced mass-loss episodes could well result in inconsistent behaviour between brightness and colour indexes, and consequently in unreliable stellar properties. Therefore, we were interested in the properties of the correlation diagrams $V$/$(B-V)$.

Contrary to YHGs (including many RSGs), Pop. I Cepheids and RR\,Lyrae stars are radially pulsating stars, and obey a number of fixed properties. Most of their light curves have steep ascending branches with respect to the descending ones. Maximum acceleration of the expansion velocity happens during this phase, and is responsible for a fast temperature rise (e.g. Pel 1978, Lub 1977, 1979). Therefore, at the same visual brightness $V$, $(B-V)$ is slightly bluer (T$_{\rm eff}$ is higher).

Our analysis of the four objects reveals a greater amount of photometric diversity than what is seen for the Cepheids and RR Lyrae stars with respect to the correlation diagrams. We noted an eye-catching difference between the $V$/$(B-V)$ diagrams of the pulsations of the hot YHG HD\,179821 (T$_{\rm eff}$\,=\,6800\,K, Arkhipova et al. 2009) and those of the cooler HR\,5171A, $\rho$\,Cas, and HR\,8752 when the latter was also as cool as its two sisters: about 5000\,K (Nieuwenhuijzen et al. 2012).

Intrinsic differences between the properties of light and colour curves between hot and cool stages of YHGs can be expected, which must be related to their evolutionary stage. Figure 1 shows that in the hot stage, HD\,179821 and HR\,8752 were crossing the Yellow Void (and HR\,8752 even evolved into the Yellow-Blue Void) along a BL evolution: the stars contracted and the mean density increased. On the contrary, the YHGs  HR\,8752, $\rho$\,Cas, and HR\,5171A in the cool stage are moving along a RL just outside the Yellow Void (Fig.\,1).
Indeed, we discovered that light curves of YHGs HD\,179821 and HR\,8752  in the hot stage after the 1973 eruption show smaller visual amplitudes (0\fm1--0\fm2) and much shorter quasi-periods (100$^{\rm d}$--200$^{\rm d}$) than the cooler YHGs. Furthermore, about  30\% of the pulsations, best documented for HD\,179821, even showed shifted colour extremes with respect to the light extremes (thus, light and colour curves do not always run in phase). Sometimes the reddest colour is reached during the ascending branch, and the bluest colour during the descent, and vice versa.

A very global conclusion is that we discovered that variable light curve properties like quasi-periods and amplitudes depend on whether the star is hot (periods are short, amplitudes are small), or cool (periods are longer, amplitudes are larger). We return to this subject in Sect.\,3. To look for the cause of the temperature inconsistency, we have to use another amplitude ratio; see the following section.

\subsection{Ratios Ampl\,$B$/Ampl\,$V$ and Ampl\,$V$/Ampl\,$(B-V)$: influence of enhanced mass loss episodes.}

Other possibilities to characterise the variability of pulsating stars are the amplitude ratios Ampl\,$B$/Ampl\,$V$ (amplitudes in $B$ are always larger than in $V$) and Ampl\,$V$/Ampl\,$(B-V)$ (for ascending and descending branches, which are mainly used here). Both ratios are a measure of the temperature change during the pulsation. The size of the amplitudes is independent of extinction and reddening corrections, and therefore so are their ratios.

--{\bf $\rho$\,Cas}: It appeared that for $\rho$\,Cas the mean ratio Ampl\,$B$/Ampl\,$V$ for tens of ascending and descending branches of pulsations (or only parts of them if there is a lack of data points, and eruptions excluded) and based on the robotic data sets 2003--2017 (by G.W.H.) is 1.6\,$\pm$\,0.3 (st.dev.). The corresponding ratio Ampl\,$V$/Ampl\,$(B-V)$\,=\,1.7\,$\pm$\,0.5 (st.dev.) It should be noted that in the 1986--2000 time interval, only $VRI$ photometry was performed. $(V-R)_{0}$ can easily be transformed into $(B-V)_{0}$ by dividing it by 0.75; see (eq.\,1) of Sheffer \& Lambert (1992).
We call these `normal' pulsations, as they are the majority. This ratio also applies to well-defined light curves of HR\,5171A and HR\,8752. We note that our designation `normal' is relative, and it only serves as a reference point.

`Abnormal' pulsations also exist, with amplitude ratios significantly larger or smaller (rare) than the value of 1.7\,$\pm$\,0.5 above. Between the two massive eruptions of $\rho$\,Cas in 1986 and 2000, many pulsations with overly large amplitude ratios occurred. That could be established because the star was monitored photometrically (1986--2000, in $UBV$ and $VRI$) and spectroscopically (1993--2002, by Lobel et al. 2003). Figure\,1 of Lobel et al. 2003 shows the $V$ and visual light curves and the V$_{\rm rad}$ curves, while the dates of the spectra and the presence of emission lines are indicated.
References used for this analysis are listed in Appendix D: 2.6.
The fact that the ratios Ampl\,$V$/Ampl\,$(B-V)$ are higher than 2.2 up to about 5, and also the relatively large variation of these ratios within this group, is a warning that some irregular phenomenon has disturbed the intrinsic photometric parameters $B$ and $V$. The height of the ratios means that the denominator is significantly smaller than for the `normal' ones, pointing to some selective absorption. Therefore, $(B-V)$ curves show overly small amplitudes, and the $(B-V)$ values are too red, causing the overly small temperature ranges.

Two such pulsations, i.e., those in 1993 ($\sim$\,JD\,24\,49300) and 1998\, ($\sim$\,JD\,24\,51065), with relatively high spectroscopic temperatures (T(Sp)\,=\,7250\,K), high maxima, and showing emission lines in the spectrum due to enhanced mass loss episodes, had ratios of 5.1 and 2.5, and 4.4 and 4.1, for ascending and descending branches, respectively (spectroscopy by Israelian et al. 1999; Lobel et al. 2003). The pulsation in 1997 ($\sim$\,JD\,24\,50575) however, that is in between the two above, without emission lines, and a with maximum rivaling the couple above, had abnormal ratios as well, 4.4 and 5.8, indicating that the absorption was still present.

Lastly, the ratio for the ascent to the minimum ($\sim$\,JD\,24\,51320) preceding the 2000-eruption max.\,1 ($\sim$\,JD\,24\,51590) is also too high: 2.5, while no obvious emission was reported. The spectroscopic temperature rose to 7600\,K, and then the brightness declined to a deep minimum, indicating that $\rho$\,Cas expanded and cooled to about $\sim$\,4400\,K.

De Jager (1998, Ch.8) suspected that pulsations during an enhanced mass-loss episode show larger light amplitudes. This appeared to be indeed the case for all eruptive maxima. However, ordinary pulsations sometimes show relatively large amplitudes during line-emission periods (e.g. 1993 and 1998), but not always like those observed in 1997 and 1998.

--{\bf HD\,179821}: Amplitude ratios were derived for most of the light curves of HD\,179821. A few detailed pulsations are shown for example in Fig.\,2 of Arkhipova et al. (2009) and in Figs.\,1, 3, and 4 of Le Coroller et al. (2003). Their light curves also include observations by ASAS, Hrivnak (2001), and Hipparcos (ESA, 1997).

Just like in the case of $\rho$\,Cas, we also identified `normal' and `abnormal' pulsations for HD\,179821 in the two time intervals 1991--1999 (JD\,24\,48500--JD\,24\,51500) and 2000--2006 (JD\,24\,51700--JD\,24\,53700) on account of the ratios. Smaller ratios than 2.2 belong to the `normal' ones, as they are the majority as well.
The mean Ampl\,$B$/Ampl\,$V$\,for HD\,179821 is\, 1.6\,$\pm$\,0.3 (st.dev., n\,=\,14), thus similar to that of $\rho$\,Cas (1.6\,$\pm$\,0.3). One peculiar pulsation, was excluded: the one of $\sim$\,JD\,24\,52500. Consequently, the mean ratio Ampl\,$V$/Ampl\,$(B-V)$\,=\,1.7\,$\pm$\,0.5 for HD\,179821 is similar to the one for $\rho$\,Cas.

At least six pulsations of HD\,179821 are abnormal (the cases with light and colour curves running out of phase were excluded, as mentioned in Sect.\,2.5., although they are abnormal as well). Indeed, often the H$\alpha$ profiles of HD\,179821 offered proof for enhanced mass-loss episodes (Tamura \& Takeuti 1991; Zacs et al. 1996); see also Arkhipova et al. (2001).
We would like to mention one concrete case with Ampl\,$V$/Ampl\,$(B-V)$\,=\,6.2, and coinciding with an enhanced mass-loss episode based on the H$\alpha$ profile, observed at JD\,24\,52793, by Sanchez Contreras et al. (2008).

--{\bf HR\,8752}: This object showed a connection between enhanced mass-loss episodes and abnormal amplitude ratios during its cool YHG stage, especially between 1976 and 1981. At the time, Smolinski et al. (1989) identified emission features in the spectrum (see Figs. 9 and 10 of de Jager 1998), while $BV$ photometry indicated overly small $(B-V)$ amplitudes (caution is called for: the number of data points was at times rather low). The photometric observations were conducted by Moffett \& Barnes 1979; Percy \& Welch 1981; Walker 1983: Arellano Ferro 1985; Zsoldos \& Olah 1985).

Only for one light curve of HR\,8752 in 1976 (thus, after the 1973 eruption) were the maximum ($\sim$\,JD\,24\,43100) and the following minimum ($\sim$\,JD\,24\,43400) properly defined by $B$ and $V$ data points, for which the amplitude ratio was relatively abnormal pointing indeed to an enhanced mass-loss episode (Smolinski et al. 1989): Ampl\,$V$/Ampl\,$(B-V)$\,=\,6.7.

In conclusion, we discovered that for YHGs, overly high Ampl\,$V$/Ampl\,$(B-V)$ values signal instable atmospheric conditions, sometimes supported by the presence of emission lines in the spectrum. The overly low $(B-V)$ amplitudes are likely caused by some gas layer with an enhanced opacity, and showing variability during the oscillations. The absorption is obviously wavelength dependent: more for $B$ (a rough estimation: of the order of 0\fm1 up to one magnitude. The higher the temperature, the higher the absorption) than for $V$, but the extinction law remains unknown. Furthermore, such a layer seems to be able to survive some successive pulsations, considering the continuation of abnormal amplitude ratios for pulsations with no noticeable emission lines.
 This appears to be the case for the time interval 1986-2000 of $\rho$\,Cas and for HD\,179821, considering the occurrence of a number of pulsations with abnormal amplitude ratios, while no emission lines were present in the spectra.

A sketch of a few fictitious pulsations in $B$ and $V$ is presented in Fig.\,D.1. of Appendix\,D:2.6.: to illustrate this cyclic behaviour with the aid of three sets of amplitudes and their ratios. More evidence for a difference between T(Sp) and T(Phot) is given in Appendix\,E:2.6.

\subsection{The temperature scales: calculations of stellar properties.}

If one would like to derive stellar properties from the photometry of many types of stars, such as the temperature (e.g. to compute M$_{\rm bol}$, the radius R and its variation), a short discussion is needed on a number of temperature calibrations for all stars based on broadband continuum photometry.

The Schmidt-Kaler (SK) calibration (1982) uses the $(B-V)_{0}$ to derive the T$_{\rm eff}$ and the BC and is valid for stars up to supergiants of luminosity class Iab only, simply because at the time there was a lack of well-calibrated super- and hypergiants.
If one uses the SK calibration of Iab stars to derive temperatures for hypergiants, they will be too high. Therefore one should use the de Jager-Nieuwenhuijzen (dJN) calibration (de Jager \& Nieuwenhuijzen 1987); see Appendix\,F:2.7.

\subsection{$\rho$\,Cas pulsations and the definition for long-term variations.}

Despite the fact that stellar parameters from YHGs derived from photometry are expected to be unreliable (Sects.\,2.4--2.6), to start with the temperature derived from $(B-V)$, it would be of most importance to start a photometric and spectroscopic monitoring program (daily) for individual pulsations.
In this way the trend of both temperatures can be studied, the purpose being to learn more about the absorption law. It would be even more rewarding if the mass loss rate were monitored as well, so that the explanation of the `cyclic absorption variation' (Appendix, Fig\,D.1.) can be verified.

As  only  high-quality photometry is required for such a program, like the multi-colour photometry of Henry (1995, 1999; see Sect.\,2.2), we selected two fine individual pulsations from the $\rho$\,Cas sequence in Fig.\,2, that is, Figs.\,4 and 5;
see Appendix\,G: 2.8. for an explanation of why spectroscopic monitoring is also indispensable.

Figure\,2 shows the 2003--2017 $BV$ sequence of quasi-periods (bright is up). The magnitudes $V$ and $B$ are relative to the comparison star HD\,223173 ($V$\,=\,5.513, $B$\,=\,7.160). Similar high-quality observations made recently by one of the current authors (EJvB) are represented by dots on the very right (see Appendix B: 2.1).

Long-term variations are the mean curves sketched through the median brightness and colour indices of all pulsations. Although already known and discussed by many researchers for decades, we are the first to suggest this definition.
The numbers placed above individual $B$ pulsations in Fig.\,2 refer to pulsations which are mentioned in this paper.
The LTV in $V$ runs almost constant over time, contrary to the one in $B$; it shows a wavy trend, with an amplitude of $\sim$\,0\fm2. This means that the individual pulsations in the maxima of the LTV ($B$) are on average bluer and hotter than in the minima.

Additionally, the LTV in $B$ also shows a significant dip (including pulsation no.\,2, depicted in Fig.\,4), lasting $\sim$\,300$^{\rm }$ d, but in $V$ it is hardly visible. A similar type of relatively fast variation, independent of the pulsations, was identified in the light variations of the other three YHGs.

\begin{figure}
 \resizebox{\hsize}{!}{\includegraphics[angle=-90]{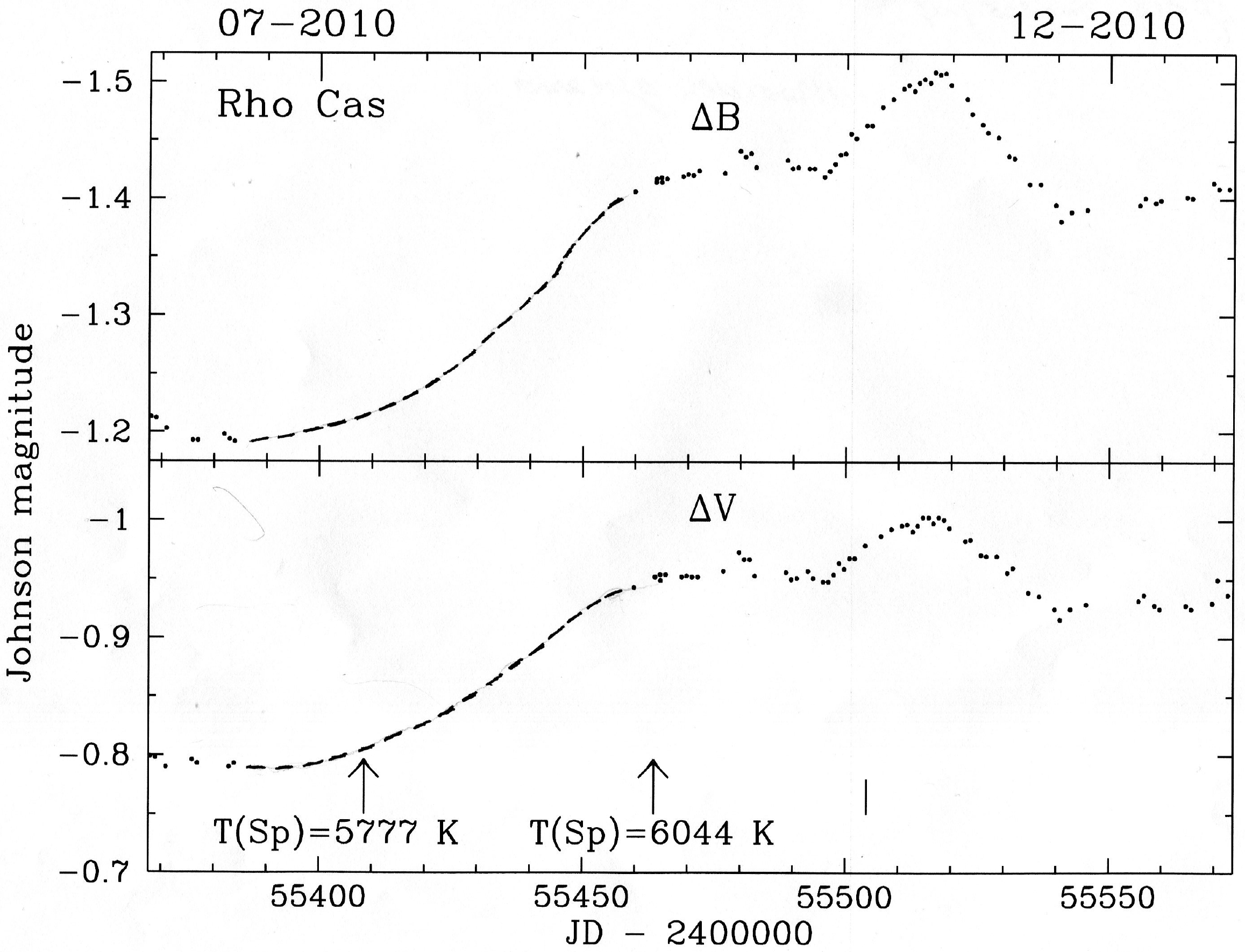}}
 \caption{ Light curve no.\,2 from Fig.\,2 of $\rho$\,Cas relative to the comparison star (in magnitudes; bright is up).  On top the light curve is remarkably dome-shaped. Dots are individual data points; the scatter is extremely small: 0\fm01--0\fm02. The dashed curve, part of the ascending branch, is regrettably not observed. The relatively large scatter of a number of data points preceding the dome near JD\,24\,55480 is due to unstable sky conditions as the reference stars showed this scatter as well. Two spectral temperatures by Klochkova et al. (2014) are indicated.}
  \label{Fig4}
 \end{figure}

Figs.\,4 and 5 show the finely detailed $V$ and $B$ light curves relative to the comparison star, with the extremely small scatter of 0\fm01--0\fm02. One can make out the slow contraction phase to maximum light and then the expansion phase (only partly) to minimum light of a very low-gravity photosphere with a huge dimension of a few astronomical units.

We selected Fig.\,4 because of two spectral temperature T(Sp) determinations, made by Klochkova et al. (2014) (JD\,24\,55409 and JD\,24\,55463). One coincides with a gap in the time-series (dashed curves) representing the minimum, and therefore interpolated magnitudes are used below, and the second one coincides with the plateau. Below, T(Sp) is compared with the calculated photometric temperatures based on the SK and dJN methods using the observed $(B-V)$ obtained at the same time.

The ascending branch in Fig.\,4 shows a plateau and a peculiar symmetrically dome-shaped maximum lasting 37$^{\rm }$ d with amplitudes of 0\fm08 and 0\fm05 in $B$ and $V$, respectively. Subsequently, the branch shows a decline almost equal to the rise, and then a plateau. After a gap in the time-series, a new pulsation started, but brighter than the previous one according to Fig.\,2.

The pulsation in Fig.\,5 (with no T(Sp) determinations), labelled no.\,3 in Fig.\,2, showed almost the same surprising light-curve morphology. The duration of the dome-shaped maximum lasts longer, namely 65$^{\rm }$ d, and the amplitudes are larger: 0\fm115 and 0\fm070 in $B$ and $V$, respectively. The ascending branch is slightly bumpy (which is intrinsic) with a timescale of weeks.

\begin{figure}
 \resizebox{\hsize}{!}{\includegraphics[angle=-90]{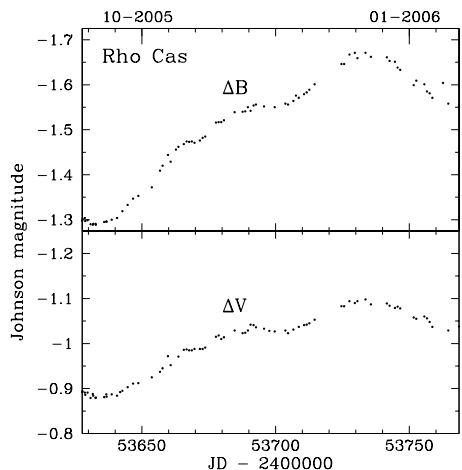}}
   \caption{Similar to Fig.\,4: but for light curve no.\,3 of Fig.\,2 of $\rho$\,Cas relative to the comparison star, also dome-shaped like in Fig.\,4. Dots are individual data points. The ascending branch shows a number of significant bumps with a timescale of a few weeks.}
   \label{Fig.5}
 \end{figure}

 A great variety of secondary features exist in other pulsations of $\rho$\,Cas, that is, no.\,5 (Fig.\,2), which has a plateau in the ascending branch lasting almost two months, and no.\,6 (Fig.\,2) shows a maximum that is regrettably interrupted by a gap in the time-series and is flat and lasts no less than four months. One can only speculate as to the causes of all these features.

Table\,2 summarises the calculated
stellar properties for the selected locations in Fig.\,4. As we assume that the plateaus represent constant temperature, radius, and so on, we have chosen the start and the end of the plateau, together with the minimum and the maximum.
The physical properties are horizontally sorted. Below the double line, we show the differences in time, $\triangle$\,JD, and the $V_{0}$ and $(B-V)_{0}$, the T(Sp) and the two calculated temperatures for Iab stars based on the SK calibration and for Ia$^{\rm +}$ stars using the dJN calibration.

Table\,3 summarises for Fig.\,4, in different temperature columns, the BC, M$_{\rm bol}$, and the radius R (in R$_{\rm \odot}$) derived from the formula:
log\,R/R$_{\rm \odot}$\,=\,8.47 - 0.2M$_{\rm bol}$ - 2log\,T$_{\rm eff}$. The symbol
$\triangle$\,R represents the decrease, or increase of R (- or +, respectively) with respect to the previous location, and the V$_{\rm rad, puls}$ is in kms$^{-1}$ (+ for contraction and - for expansion) between two successive locations (vertically sorted). We note that they represent the mean velocity between the two locations. We used the accepted distance of 3.1\,kpc.
 As YHGs are in general non-radial pulsating stars (Sect.\,2.2.), the $\triangle$\,R, and V$_{\rm rad,puls}$ are lower limits.

\begin{table}
\caption{Extinction free visual brightness, colour indices and three types of temperature determinations for four instances in time in the ascending branch (two for the plateau: the first for the beginning and the second for the end of the plateau) of a pulsation of $\rho$\,Cas shown in Fig.\,4. We note that T(SK) and T(dJN) are based on the $(B-V)_{0}$, therefore, they are photometric temperatures T(Phot).}
\label{table:2}
\centering
\begin {tabular}{c|c|c|c|c}
\hline\hline
Location & min & plateau & plateau & max \\
\hline\hline
JD-- & 5409 & 5464 & 5496 & 5515 \\
245\,0000 & & & & \\
\hline\hline
$\triangle$\,JD & -- & 55\,d & -- & 19\,d \\
$V_{0}$ & 3.28 & 3.12 & 3.13 & 3.07  \\
$(B-V)_{0}$ & 0.79 & 0.73 & 0.72 & 0.70 \\
T(Sp) & 5777\,K & 6044\,K      \\
T(SK)  & 5450\,K & 5640\,K & 5660\,K & 5710\,K  \\
T(dJN) & 4890\,K & 4980\,K & 5000\,K & 5040\,K \\
\hline
\end{tabular}
\end{table}

\begin{table}
 \caption{BC, M$_{\rm bol}$, and radius R for four locations in the ascending branch (two for the plateau: the first for the beginning and the second for the end) of the pulsation of $\rho$\,Cas shown in Fig.\,4, the radius variation $\triangle$\,R and contraction/expansion velocity $V_{\rm rad, puls}$. The latter represents the mean velocity between the selected locations. We note that they represent pure radial velocities of the continuum radiating layer, and are independent of the space velocity. }
 \label{table:3}
 \centering
 \begin{tabular}{c|c|c|c}
  \hline\hline
  & T(Sp) & T(SK) & T(dJN) \\
  & {\bf min} & {\bf min} & {\bf min} \\
 \hline\hline
 & 5777\,K & 5450\,K & 4890\,K  \\
 \hline
  BC & -0.13 & -0.16 & -0.20 \\
  M$_{\rm bol}$ & -9.31 & -9.35 & -9.39 \\
  R &     668\,R$_{\rm \odot}$ & 779\,R$_{\rm \odot}$  & 970\,R$_{\rm \odot}$ \\
 \hline\hline
  & T(Sp) & T(SK) & T(dJN) \\
  & {\bf plateau} & {\bf plateau} & {\bf plateau} \\
 \hline\hline
 & 6044\,K & 5640\,K & 4980\,K \\
 \hline
 BC & -0.10 & -0.14 & -0.19 \\
 M$_{\rm bol}$ & -9.43 & -9.48 & -9.53 \\
 R & 636\,R$_{\rm \odot}$ & 747\,R$_{\rm \odot}$ & 981\,R$_{\rm \odot}$ \\
$\triangle$\,R & -22\,R$_{\rm \odot}$ & -7\,R$_{\rm \odot}$ & +27\,R$_{\rm \odot}$ \\
 V$_{\rm rad, puls}$ & +3.2\,kms$^{-1}$ & +1.0\,kms$^{-1}$ & -4.0\,kms$^{-1}$ \\
 \hline\hline
 & T(Sp) & T(SK) & T(dJN) \\
 & {\bf plateau} & {\bf plateau} & {\bf plateau} \\
 \hline\hline
  &  & 5660\,K & 5000\,K \\
  BC & & -014 & -0.18  \\
 M$_{\rm bol} $ & & -9.46 & -9.51  \\
 R &  & 735\,R$_{\rm \odot}$ & 964\,R$_{\rm \odot}$ \\
 \hline\hline
  & & T(SK) & T(dJN) \\
  &  & {\bf max} & {\bf max} \\
 \hline\hline
  & & 5710\,K & 5040\,K \\
 \hline
 BC & & -0.13 & -0.17  \\
 M$_{\rm bol}$ &  & -9.52 & -9.56  \\
R & & 742\,R$_{\rm \odot}$ & 971\,R$_{\rm \odot}$ \\
$\triangle$\,R &  & +7\,R$_{\rm \odot}$ & +7\,R$_{\rm \odot}$  \\
V$_{\rm rad, puls}$ &  & -3.0\,kms$^{-1}$ & -3.0\,kms$^{-1}$ \\
\hline
\end{tabular}
\end{table}

 The summary below lists the average differences between the calculated temperatures and radii based on different temperature scales.
The average differences in T and R according to Tables\,2 and 3 and based on two temperature scales are:
T(Sp)--T(dJN)\,$\sim$\,990\,K,  R(Sp)--R(dJN)\,$\sim$\,-300\,R$_{\rm \odot}$.

According to Klochkova (2014) and Klochkova et al. (2014), the uncertainties in T(Sp), based on ratios of selected spectral lines being sensitive temperature indicators, are in the range of 40\,--160\,K only. Considering the temperature differences we report, as shown in Tables\,2 and 3, these differences are negligible. Therefore, the differences between T(Sp) and T(Phot) (=\,T(dJN) and T(SK)) in Tables\,2 and 3 are due to the effect we discovered, and for which we offer an explanation in Appendix\,Fig.\,D.1.

Obviously, the observed $(B-V)$ is too red, and is therefore inappropriate for deriving accurate temperatures and all other stellar properties.
For example, the trend of the calculated photometric radii at the four locations in Fig.\,4, especially for the YHG calibration, are incorrect. They clearly run opposite to radius variations of YHGs, and should decrease towards maximum brightness (=\,contraction, plus sign) instead of increasing. Its ratio Ampl\,$V$/Ampl\,$(B-V)$\,=\,2.3, meaning that the $B$ amplitude with respect to $V$ amplitude is indeed too small (Sect.\,2.6.). This suggests that absorption increases with temperature: at maximum light the absorption of $B$ increased with respect to the absorption in $V$.

The V$_{\rm rad}$ value of 3.2\,kms$^{-1}$ (second column), which is based on spectroscopy and should be multiplied by 1.4 to correct for projection effects, and so on, could become of the right order if the overly low radius variation (for which the correction is unknown) is also taken into account: Lobel et al. 2003: 5--10\,kms$^{-1}$ and Klochkova et al. 2014: 7\,kms$^{-1}$.

The above is in support of the existence of a cyclic absorption behaviour (Appendix\,Fig.D.1.): as a consequence, the photometry of $\rho$\,Cas (and of all YHGs) does not represent a normal YHG photosphere. Stellar properties based on photometric data weakened by absorption with an unknown absorption law are unreliable. Again, we estimated that the absorption in $B$ is of the order of 0\fm1 up to a few times this value; in $V$ the absorption is presumably about a factor two smaller, but that is very uncertain. The higher the mass-loss rate and the higher the temperature, the higher the absorption.

The photometric parameters of the pulsation in Fig.\,5 (no spectral temperatures known), of which the amplitude ratio is 1.2, and thus `normal', are less inconsistent, yet the calculated temperatures (dJN) are still too low by hundreds of degrees with respect to the T(Sp) in Fig.\.4, while both pulsations do not differ much with respect to the medium brightness. Although the photometry of the other YHGs are less accurate, our conclusions were similar. Therefore, we suspect that a high-opacity layer is always present in the atmospheres of YHGs, and its absorption capacity depends on the density of that layer (higher by enhanced mass loss) and on the temperature (higher by a temperature rise), and vice versa.

The magnitudes of $\rho$\,Cas, used in Tables\,2 and 3 offer a chance to calculate errors due to scatter as small as $\pm$\,0\fm01. The result is approximately $\pm$\,15\,K and $\pm$\,10\,R$_{\rm \odot}$.
The relatively large size of these errors demonstrates the extreme sensitivity  of the calculated stellar properties to `very small' photometric errors.
Our message is that this analysis underlines the importance of a long-lasting photometric and spectroscopic monitoring all over the world; see Appendix\,G: 2.8.

\subsection{Proposed distance reductions for $\rho$\,Cas and HR\,5171A.}
Below we propose shorter (conventional) distances for $\rho$\,Cas and HR\,5171A. We ignore the distances based on parallaxes by the satellites Hipparcos and Gaia (Sect.\,1).

The size of the radii derived from the spectroscopic and radial velocity
monitoring campaign during the eruption in the year 2000  motivated us to question whether the accepted distance of $\rho$ Cas by Zsoldos \& Percy (1991) of 3.1\,kpc\,$\pm$\,0.5\,kpc is too large. We note that the sizes of the radii obtained in this way were derived independently of the distance by Lobel et al. (2003). They estimated that the extreme dimensions of the stellar radius at the eruption maximum and at the deep minimum brightness were $\sim$\,400\,R$_{\rm \odot}$ and $\sim$\,1000\,R$_{\rm \odot}$, respectively.

The calculated mean radius of an average pulsation with a normal amplitude in $V$ of for example  0\fm21, is $\sim$\,960\,R$_{\rm \odot}$ (dJN), which is too close to the radius for the deep eruption (photometric) minimum (when the star has reached its maximum size) to be plausible. Radii based on the SK calibration are about 200\,R$_{\rm \odot}$ smaller, but this is definitely only valid for SGs of type Iab. If de Jager \& Nieuwenhuijzen (1987) were found to have assumed overly extreme properties for hypergiants relative to the Iab SGs, then a reduction of the radius difference above from 200\,R$_{\rm \odot}$ to say 100\,R$_{\rm \odot}$ would yield a radius of 860\,R$_{\rm \odot}$ (instead of the above 960\,$_{\rm \odot}$), which would still be too high to be credible. Therefore, the error on the dJN radius cannot be much more than that on the SK radius.

Other options are to decrease the interstellar extinction from 1\fm44 (Table\,1) to 1\fm0 for example, and the reddening from 0\fm45 to about 0\fm30, but these values are not likely. Our choice of 2.5\,kpc\,$\pm$\,0.3\,kpc is rather arbitrary. The M$_{\rm bol}$ becomes about 0\fm45 less negative (luminosity lower), and the calculated radii are reduced by a factor of $\sim$\,1.2: $\sim$\,770\,R$_{\rm \odot}$, as well as the radius variations and the radial pulsation velocities. For us, these values are more acceptable with respect to the derived extreme dimensions above.
This case serves as a reminder that our attempts to obtain more precise information on YHGs are seriously hampered as long as distances remain uncertain, not only for $\rho$\,Cas, but also for HR\,5171A below, and HD\,179821.

The accepted distance of HR\,5171A is 3.6\,kpc, assuming that it belongs to Gum48d and that HR\,5171B, at a distance of 10" is a nearby companion: optically or physically (Humphreys et al. 1971; Schuster 2007; Schuster et al. 2006). However, subsequently M$_{\rm bol}$\,=\,-10 -- -11 depending on the reddening (1\fm0--1\fm4) and reddening law used (3.1--3.5), creates a problem, as this is much too bright for a hypergiant. As far back as 2013 one of us (HN) emphasised that the accepted distance is too large. As a result, the radius would then also be much too large: between 6 and 12\,AU. We think that this is too big, until new outcomes are undisputable (van Genderen et al. 2015).

A distance of between 1 and 2\,kpc for example, would entail a luminosity that is more in accordance with its Ia$^{+}$ hypergiant character, for example with M$_{\rm bol}$ lying between -8.7 and -9.6 (Sect.\,1) and a radius between 3 and 5\,AU. We do not think that HR\,5171A is a RSG as advocated by Wittkovski et al. (2017a); see Sect.\,2.2. HR\,5171A shows too many similarities with $\rho$\,Cas for example, like its eruptive activity, and above all, the spectra of the Big Three are almost identical. However, the H$\alpha$ line of HR\,8752 strongly differs with respect to the shape because of the permanent presence of a strong emission line.
Thus, the physical properties of the winds of the Big Three are in many respects identical (Lobel et al. 2015); in any case, RSGs do not show eruptive events.

A portion of the light curve between JD\,24\,47200 and JD\,24\,48800 (1988--1992) was spectroscopically monitored, resulting in a well-covered V$_{\rm rad}$ curve with an amplitude of 10\,kms$^{-1}$ (Lobel et al. 2015). This correlated very well with the $V$ magnitudes between the maxima of pulsations\,16 and 17 (numbered according Fig.\,1 in Paper\,I), indicating a steady contraction towards maximum brightness and the reverse to the brightness minimum.
We note that at the time HR\,5171A had reached its faintest brightness: $V$\,$\sim$\,7\fm0 and reddest colour so far: $(B-V)$\,$\sim$\,2.6 (Figs.\,13 and 14).
Appendix\,H.2.9. presents a description of a method which is in principle suitable for deriving the distance of a pulsating star with the aid of the calculated V$_{\rm rad puls}$ and the spectroscopically obtained V$_{\rm rad}$, but $B$ and $V$ magnitudes of the pulsation cycle should be undisturbed.

Another argument favouring a much shorter distance than 3.6\,kpc is based on the fact that at this distance, the position of HR\,5171A would be below the Humphreys-Davidson limit on the theoretical HR diagram (shown in Fig.\,6 of Paper\,I) instead of far above it. At below 3.6 kpc, the distance with respect to the semi-empirical P\,=\,constant line for $\sim$\,500 d (the mean quasi-period of HR\,5171A between the 1960s and 1990s; van Leeuwen et al. 1998) would be more in accordance with the positions of the other variable supergiants.

Further, the variability of the pulsation quantity Q\,=\,P$\rho^{-1/2}$ appeared to be compatible with stability studies of stellar models and was first described by Maeder \& Rufener (1972). Based on that study, Burki (1978) derived a formula between the dependency of the quasi-period P on the mass M/M$_{\rm \odot}$, M$_{\rm bol}$ and on T$_{\rm eff}$.
Applying in his Eq.(5) the approximate but plausible input parameters 25\,M/M$_{\rm \odot}$, -8.8, and 4300\,K, respectively, the quasi-period becomes P = 473$^{\rm d}$, that is, roughly of the same order as the observed mean quasi-periods (see Fig.\,13). By varying the input parameters by trial and error, say by $\pm$\,5\,M/M$_{\rm \odot}$, $\pm$\,0\fm2 and $\pm$\,200\,K, respectively, in either directions, results are of the same order.

A third argument for a shorter distance refers to the energy budget of the Gum48d nebula.
Schuster (2007) considered HR\,5171B as the single central engine of the associated HII region RCW80, just like Karr et al. (2009). Based on calculations on the present energy budget, the latter concluded that the nebula only needed one single ionizing O-type star, that is, the present B-type star HR\,5171B, only a few million years ago still on the main sequence, just like HR\,5171A at the same time. In other words, for the present energy budget of Gum48d, the presence of HR\,5171A as a member of Gum48a is superfluous.
If HR\,5171A is indeed a foreground star, a new place of birth should be found.

On the contrary, Humphreys et al. (1971) and Schuster (2007) offered a number of arguments based on various independent techniques and physical considerations favouring the larger distance. Therefore, the distance remains debatable for the time being.

In conclusion, if our preferred  distance of 1.5\,kpc\,$\pm$\,0.5\,kpc were found to be correct, it would dethrone HR\,5171A as one of the biggest stars known (Chesneau et al. 2014; Wittkovski et al. 2017c).

This distance reduction would also reduce the flux of the reported Blue Luminescence by PAH molecules in the 1970s (van Genderen et al. 2015) by a factor $\sim$\,6.
This is because the excesses were measured relative to the stellar flux in the $L$ channel and would stay unaffected for both scenarios (1) and (2): i.e. the source of the excess lies in Gum48d, or in the outer envelope of HR\,5171A, respectively.
HR\,5171A possesses an optically thick extended molecular and dust envelope at about 1.5 stellar radii (Wittkovski et al. 2017a). According to studies by Gorlova et al. (2009) and Oudmaijer \& de Wit (2013) a dense optically thick layer, formed by the wind just above the photosphere is able to shield the stellar radiation as well. Thus, there are ample possibilities to locate neutral PAH molecules to become excited to the upper electronic state by high energetic photons (3.5--5eV, A.N. Witt 2013, priv.comm). There after, these molecules recombine and emit within a near-UV band (coinciding with the $L$ band of the Walraven $VBLUW$ photometric system) radiation, called blue luminescence, discovered by Vijh et al. (2004) in the Red Rectangle nebula.

\section{The long-term variations}

 \subsection{Light and colour curves, and trends of light amplitudes and quasi-periods.}
 In this section we outline the photometric properties of LTVs (on which the pulsations are superimposed and that was defined in Sect.\,2.8.) exhibited by the four selected objects. Our interest is also focussed on their possible influence on pulsations and the occurrence of eruptions which has never been explored before.

Light and colour curves of LTVs often appear whimsical with timescales of hundreds up to a few thousand days. The evolutionary models usually reveal straight tracks on the HR diagram. Yet, we wondered whether LTVs represent in fact the evolutionary tracks, which once transformed into luminosity and temperature might not
be straight tracks on the HR diagram.
 Whether straight or irregular, the subsequent zigzag movements of the YHGs on the HR diagram are supposed to be part of the need to loose enough mass for the final course to the blue (Sect.\,1, Fig.\,1).

 These LTVs have already been noted and discussed by numerous researchers and observers in the past for HR\,5171A  (see for references Appendix\,I:3.1.), but generally they were reticent about a possible evolutionary origin, contrary to our present suspicions based on the thorough study of HR\,8752 from 1850 until 2005, including the eruptive episode from 1973 by Nieuwenhuijzen et al. (2012). These latter authors convincingly showed that the long-term variations of HR\,8752 shown in our Figs.\,6 and 7 are evolutionary tracks, or loops: a red one until the eruptive episode in 1973, and then a blue one (Figs.\,5, 6, 10, etc. in Nieuwenhuijzen et al. 2012). These evolutionary loops (RL and BL) represent the red and blue tracks (no.\,3 and no.\,4) of the models of Meynet et al. (1994), respectively.

The most obvious conclusion from the study by Nieuwenhuijzen et al. is that the LTVs of the other three objects should have something to do with evolution as well. To prove that conjecture, we scrutinized the literature for scattered photometric observations of pulsations from which the LTV  at the time could be derived. More or less complete light and colour curves were put together (thus curves sketched through the median brightness and colour). As this has never been done before, and because of the importance of these curves we show them in their entirety in Figs.\,6--14. In Tables\,4 and 5 of Sect.\,3.2 we tabulate the characteristic zigzag changes of brightness and colours, still referred to as `evolutionary modes', and their timescales.

Large empty intervals in time series, portions with bad sampling, and the presence of eruptive phases decrease the reliability of sketched LTVs, but generally they cannot be grossly in error. In the case of scattered historic visual or photographic observations of stars dating back to 1850, like in the case of HR\,8752 (Luck 1975; Arellano Ferro 1985; Zsoldos 1986a, b; Nieuwenhuijzen et al. 2012), we connected individual data points and the averages of small clusters of data points.

Because of their importance these curves merit careful description.

{\bf --HR\,8752:} Figures\,6 and 7 depict the LTVs for HR\,8752 in $V$, $B,$ and $U$, and $(B-V)$ and $(U-B)$, respectively, for the time interval 1941--1994.

Based on a collection of numerous historical data since 1840, and modern photometric data of HR\,8752, Nieuwenhuijzen et al. (2012) concluded that HR\,8752 evolved through an increasing reddening and cool episode in the 1960s and 1970s. It is noteworthy that the corresponding red minimum in our Fig.\,7 is also obvious in the log\,T$_{\rm eff}$/JD diagram of Fig.\,10 of Nieuwenhuijzen et al. (2012). These latter authors suggested that a massive eruption should have happened around 1973, although photometry was absent.

Their assumption was based on spectroscopic observations made at the time.
Luck (1975) derived from spectral scans a temperature of 4000\,K in August 1973, and $(B-V)$\,=\,1.78 in November 1974 but alas, by the scarcity of photometric $B$ and $V$ magnitudes, we could not calculate T(Phot), despite the expected amplitude of $\sim$\,1\fm0. This can be explained by the fact that the gaps in the time-series are much larger than the duration of an eruption: 400$^{\rm d}$--700$^{\rm d}$.

Additionally, we found out that light amplitudes and quasi-periods of HR\,8752 since 1976 decreased linearly until 1993 to about 0\fm05--0\fm1 and 100$^{\rm d}$--150$^{\rm d}$, respectively. This is a consequence of the continuous contraction of the star and the increase of its atmospheric density.

\begin{figure}
  \resizebox{\hsize}{!}{\includegraphics[angle=-90]{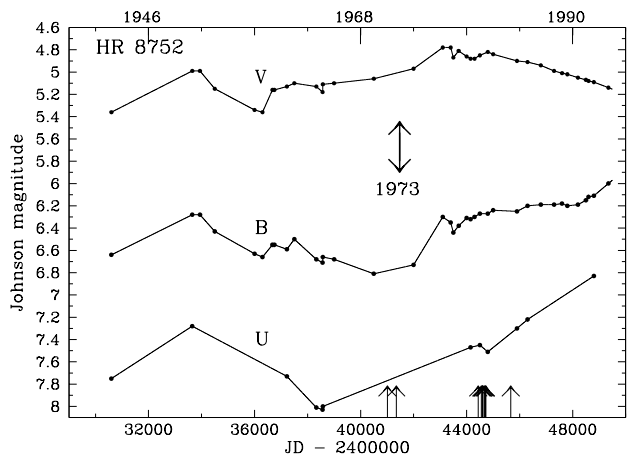}}
  \caption{Long-term variables of HR\,8752 in $V$, $B,$ and $U$ in  1942--1994. The arrows at the bottom indicate dates of spectroscopic observations by Smolinski et al.
  (1989), taken in 1969 and 1970, and between 1979 and 1982; see also
    de Jager (1998, Figs.\,9 and 10 in Sect.\,7). These spectroscopic observations signalled the presence of an enhanced mass-loss episode and a massive eruption in 1973 (arrow), which was not photometrically recorded
  (Nieuwenhuijzen et al. 2012). We note that after 1979 ($\sim$\,JD\,24\,44000) only $V$ declined; see text.}
  \label{Fig6}
  \end{figure}

\begin{figure}
 \resizebox{\hsize}{!}{\includegraphics[angle=-90]{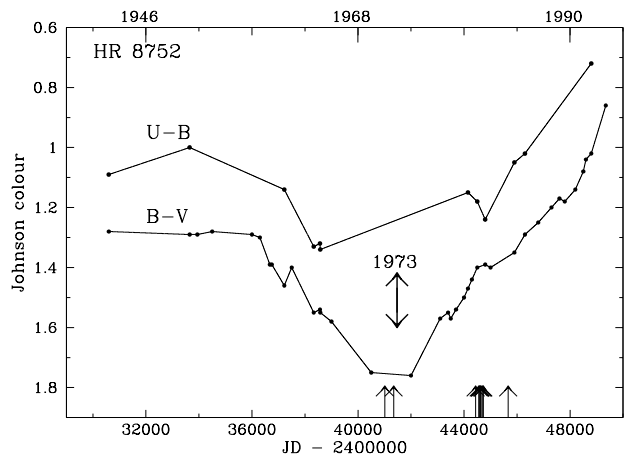}}
  \caption{Similar to Fig.\,6, but now for the LTV of $(B-V)$ and $(U-B)$ of HR\,8752, 1942--1994. We note that after JD\,24\,36000 (1957) the star suddenly reddened during the next $\sim$\,16\,y by $\sim$\,0\fm5 until the eruption in 1973. The spectral study by Luck (1975) revealed a temperature of 4000\,K at JD\,24\,41910, August\,1973.   }
   \label{Fig7}
 \end{figure}

  \begin{figure}
   \resizebox{\hsize}{!}{\includegraphics[angle=-90]{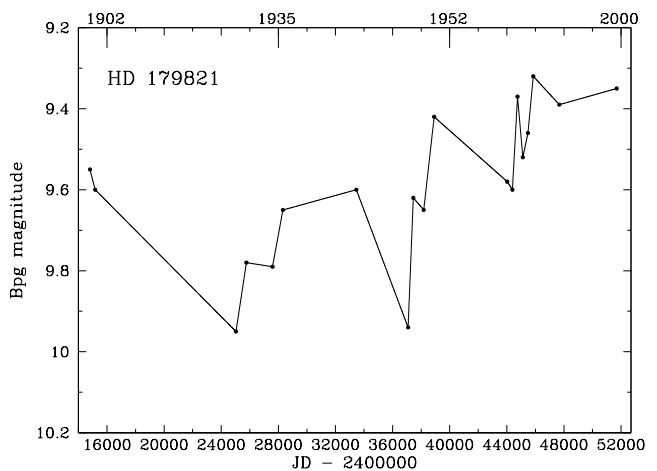}}
   \caption{Photographic LTV ($B_{\rm pg}$) of HD\,179821 between 1899 and 1989 (Arkhipova et al. 2001). We note the deep minima between JD\,24\,24000 and JD\,24\,38000 ($\sim$\,1925 and $\sim$\,1960) indicating a very red and cool episode, favourable for an eruption; see text. HR\,8752 showed the same type of photometric history before its blue loop evolution, in which an eruptive episode occurred (1973) according to spectroscopic observations (Figs.\,6 and 7). }
   \label{Fig8}
 \end{figure}

\begin{figure}
   \resizebox{\hsize}{!}{\includegraphics{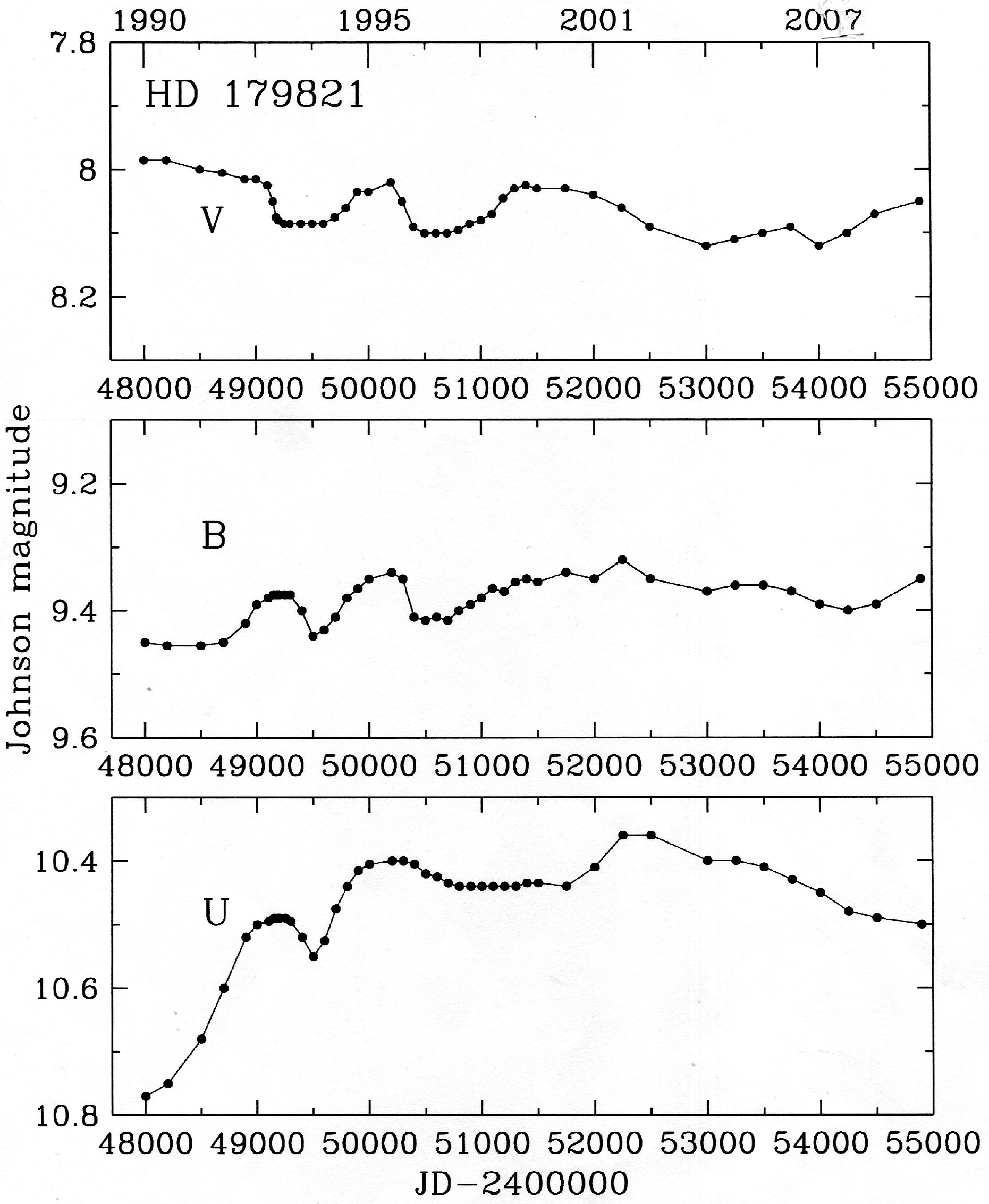}}
  \caption{Splendid coverage of the LTVs in $V$, $B$, and $U$ of HD\,179821, between 1990 and 2009 (based on observations by Arkhipova et al. 2001, 2009; Le Coroller et al. 2003;
  and Hrivnak et al. 2001). We note that dips and bumps in $V$, $B,$ and $U$ do not vary in concert. It appears that the LTVs in $JK$ of Arkhipova et al. 2009 and Hrivnak et al. 1989 (see Fig.\,9 in Arkhipova et al. 2009) mainly mimic the $V$ trend, and more or less the bumps and dips between JD\,24\,49000 and JD\,24\,50500. We witness here the presumable evolution of HD\,179821 along a blue loop in the direction of the Yellow Void (just like HR\,8752 did after 1990) ending around JD\,24\,53000 (2004), after which HD\,179821 moved back on a red loop, see text; see Fig.\,10.}
  \label{Fig9}
  \end{figure}

 \begin{figure}
  \resizebox{\hsize}{!}{\includegraphics[angle=-90]{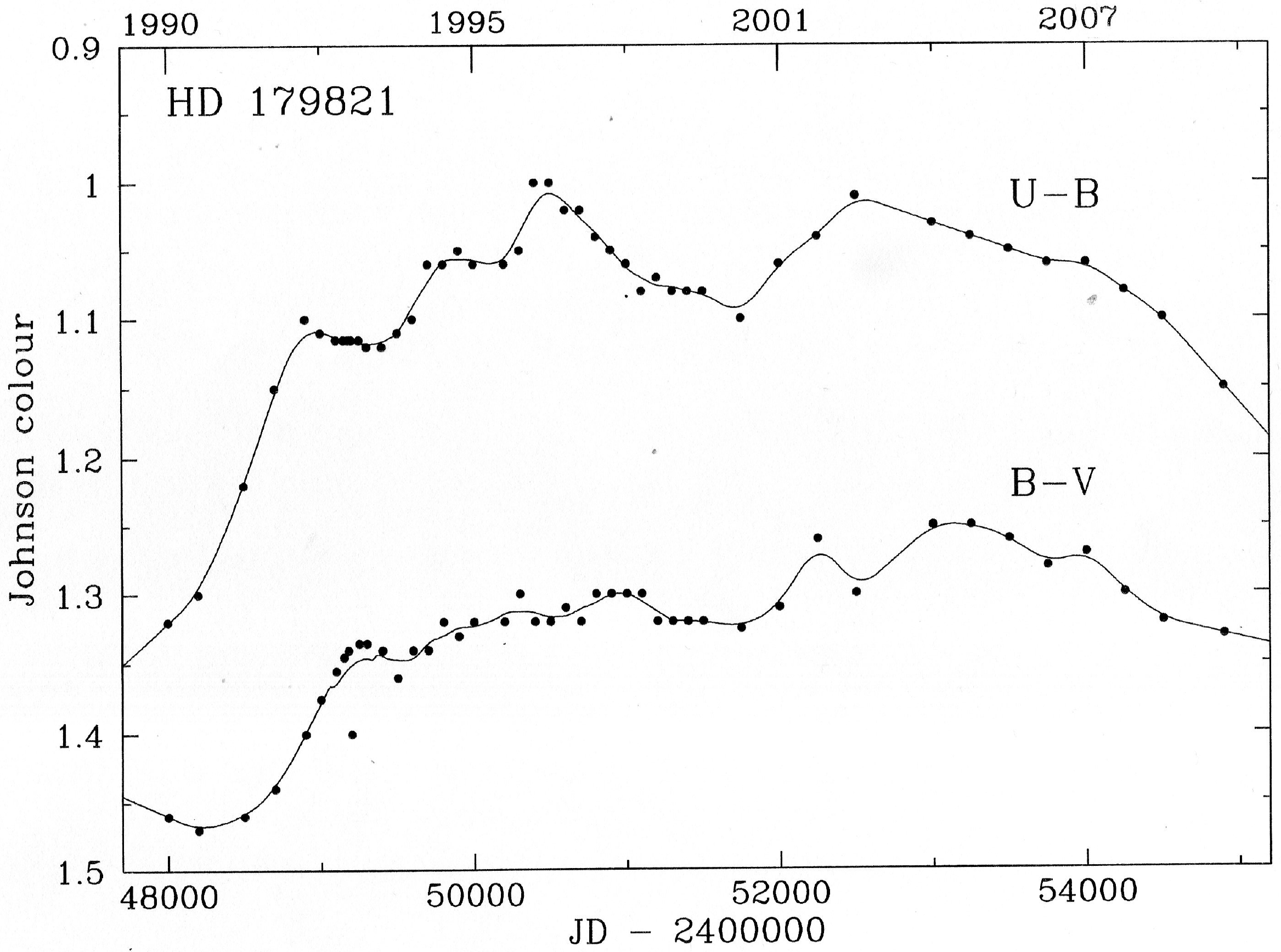}}
  \caption{Same as Fig.\,9 but for $(B-V)$ and $(U-B)$ of HD\,179821 between 1990 and 2010. The bluest value reached during the blue loop evolution is $(B-V)$\,=\,1.25, thus, $(B-V)_{0}$\,=\,0.55 (extinctions in Table\,1) happened at JD\,24\,53000 (2004). The mean spectroscopic temperature T(Sp) at the time according to Arkhipova et al. (2009) is 6800\,K\,$\pm$\,50\,K. After 2004 the star moved back on a red loop, see text.}
  \label{Fig10}
  \end{figure}

{\bf --HD\,179821:} The first record of an LTV of HD\,179821 started in 1899 (JD\,24\,16000) with a photographic $B_{\rm pg}$ dataset until 1989 (collected by Arkhipova et al. 2001).
Individual data points, or small concentrations of data points are connected by a line (Fig.\,8). This LTV is of course less reliable between 1899 and 1989, yet it gives a good impression of the  instability of the star at the time and its cool phase.
 Unfortunately, no details of a possible eruption were observed, but we consider it not unlikely that one occurred somewhere between 1925 and 1960.
Our assembled data show on average a gradual rise in light amplitude of $\sim$\,0\fm5. There is some overlap with modern observations by means of two data points which precisely match the modern $B$ magnitude sequence made photoelectrically until about 1989 ($\sim$\,JD\,24\,48000, Figs.\,9 and 10). Therefore, there is no significant reason to mistrust the reliability of the observations dating back to 1899. Thereafter, the photometric instability from the visual to the UV of the LTVs was very small: 0\fm1--0\fm2, until 2009 (JD\,24\,53000).

It appears that the $B_{\rm pg}$ was 9.5 in 1899 ($\sim$\,JD\,24\,15000), declined to a deep minimum with $B_{\rm pg}$\,10 around 1925 ($\sim$\,JD\,24\,24000), and was at a second deep minimum around 1960 ($\sim$\,JD\,24\,37000). The $B$ in Fig.\,9 gradually rose to a magnitude of 9.4--9.3 in 2009 ($\sim$\,JD\,24\,55000), showing small fluctuations, at most by $\pm$\,0\fm1, just like in $V$.

The modern portion of the LTV in $U$ (Fig.\,9) from JD\,24\,48000 until JD\,24\,52500 (1990--2003) shows that the star was still rising in brightness by 0\fm4.

The gradual blueing of $(B-V)$ in Fig.\,10 between JD\,24\,48000 (1990) and JD\,24\,53000 (2004) and the subsequent reddening until 2009  is of crucial importance for the interpretation of the evolutionary state of HD\,179821 below. This reddening trend is supported by the new $UBV$ photometry until 2017 by Ikonnikova et al. (2018). If HD\,179821 is indeed a YHG an eruptive episode can be expected within a few decades.

Noteworthy are two bumps and dips in the LTVs for $V$, $B,$ and $U$ in Fig.\,9 between JD\,24\,49000 and JD\,24\,50500. Sometimes $UB$ and $V$ run in opposite directions. The LTVs in $JK$ shown in Fig.\,6 of Arkhipova et al. (2009) mimic the one in $V$. The same is more or less the case for the bumps and dips.

\begin{figure}
   \resizebox{\hsize}{!}{\includegraphics[angle=-90]{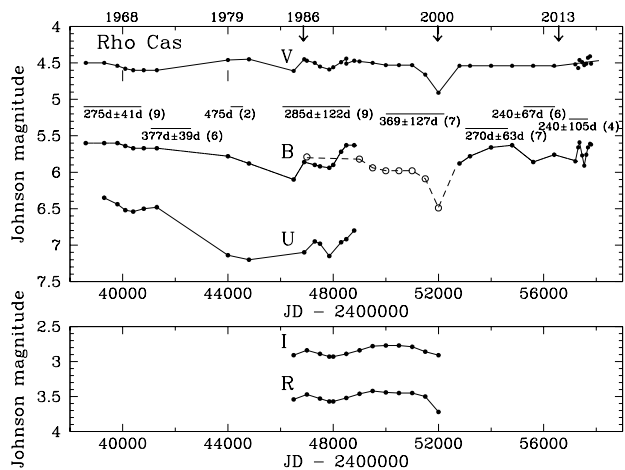}}
   \caption{Long-term variations in the brightness $UBVRI$ (Johnson system) of $\rho$\,Cas between 1965 and 2017. The arrows and the dates at the top mark eruptive episodes. Data points at JD\,$\sim$\,24\,58000 are observations by EJvB. Between the $V$ and $B$ light curves the mean quasi-periods are shown with their $\sigma$ for some selected time
  intervals. Their mean durations are represented by the length of the bars. The number of cycles used are bracketed.
  A correlation between the trend of the mean quasi-periods and the $(B-V)$ is obvious: they increase and decrease when the $(B-V)$ becomes red and blue, respectively (while $V$ is almost constant, $B$ shows a cyclic fainting and brightening before and after an eruption).
  The eight $B$-data points represented by circles are based on a $(V-R)$/$(B-V)$ relation
 when no $B$ observations were made. }
  \label{Fig 11}
  \end{figure}

 \begin{figure}
   \resizebox{\hsize}{!}{\includegraphics[angle=-90]{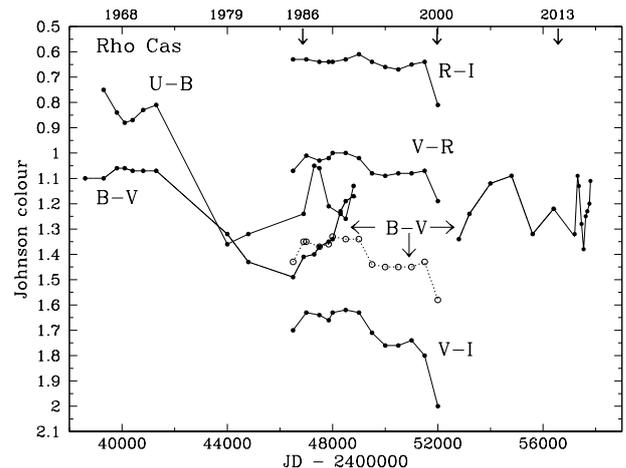}}
    \caption{Similar to Fig.\,11 but for the five colour indices of the $UBVRI$ (Johnson system) of $\rho$\,Cas between 1965 and 2017. The $(B-V)$ data points
  represented by circles are based on a $(V-R)$/$(B-V)$ relation when no $B$ observations were made (Fig.\,11).
  The arrows and the dates at the top mark eruptive episodes. We note that eruptions preferably take place when the LTV of $(B-V)$ is relatively red, therefore when the star resides in the red part on the HR diagram.}
    \label{Fig 12}
    \end{figure}

 {\bf --$\rho$\,Cas :} Figure\,11 shows the LTVs of $\rho$\,Cas for the $UBVRI$ data between 1965 and 2015. The arrows at the top indicate observed eruptive episodes. The mean quasi-periods with standard deviations and the number of cycles used (bracketed) are given in the second panel. The selected time intervals are indicated by the length of the bar.
 Figure\,12 shows the LTVs for the colour indices $(B-V)$, $(U-B)$, $(V-R),$ and $(V-I),$  the magnitude scale of which is about three times larger than in Fig.\,11. The LTVs are based on the observations by many observers (Sect.\,2.1.), but the majority are by Henry (1995, 1999). The preference of eruptions to occur when the star is relatively faint and red is obvious.

The correlation between the trends of the quasi-periods and the $(B-V)$ is obvious and depends on the type of the evolutionary track: increasing and decreasing during RL and BL evolutions, respectively (Figs.\,11 and 12), just like HR\,8752.

{\bf --HR\,5171A :} For HR\,5171A we had the disposition of photographic $B$ magnitudes (Johnson system) to reconstruct its photometric history from 1900 until 1950 and between 1978 and
  1989. They were obtained from Harvard plates of which the scanning was made within the framework of the DASCH scanning project. The measurements were kindly put at
 our disposal by Dr. Josh Grindlay, head of this project, before the official release of the data.
 The position of HR\,5171A was close to the edge of the plate. Probable errors are 0\fm2--0\fm3. It should be noted that the Harvard magnitudes include the nearby blue optical
  companion HR\,5171B (10$\arcsec$), which is likely not a physical companion, but a member of Gum48d, contrary to HR\,5171A, at least if its proposed shorter distance in Sect.\,2.9. is correct. Its contribution to the total brightness is, depending on the brightness of HR\,5171A, smaller than $\sim$~0\fm25.
  According to Stickland \& Harmer (1978), HR\,8752  also has a nearby B-type companion, but that one is a physical companion; see also Nieuwenhuijzen et al. (2012, Sect.\,1.1.

  The brightness limits ($B$\,=\,8\fm4--9\fm8) and the mean magnitude (9\fm1) of the data points in the 1900--1950 set which is not shown in the present paper appear to be similar to the modern ones after 1952 (Fig.\,1 in Paper\,I, and Fig.\, 3 in Chesneau et al. 2014). These papers refer to the original sources like those of Harvey 1972; Humphreys et al. 1971; Dean 1980; from ASAS-3 by Pojmanski 2002; from AAVSO amongst others by Otero; and from the Long-Term Photometry of Variables (LTPV) group: Sterken 1983, Manfroid et al. 1991, Sterken et al. 1993; from Hipparcos discussed by van Leeuwen et al. 1998. The Harvard $B$ data collected in the time interval 1978--1989 match the photoelectric data satisfactory. Our conclusion is that HR\,5171A did not alter its variability pattern during the last 120\,yrs.

  Figures\,13 and 14 represent the LTVs of HR\,5171A in $B$ and $V$, and in $(B-V)$, respectively, between 1953 and 2018: the black lines connecting the black dots. These latter figures are mainly based on the papers by van Genderen (1979 and 1992 Fig.\,1) and Chesneau et al. (2014) discussing the photometric history of HR\,5171A. The meaning of the other curves in Fig.\,14 is explained in Sect.\,3.4.

  Between 1994 and 2012 observations were made only by visual observers, but the means, represented by the dashed oscillating curves fit the sketched LTV in $V$ reasonably well. However, some systematic deviations can be attributed to the fact that for the construction of part of the LTV (between JD\,24\,50900 and JD\,24\,55500), we relied mostly on those made by Otero (see Chesneau et al. 2014) and on the All Sky Automated Survey-3 (ASAS) data (Pojmanski 2002; see Fig.\,3 in Chesneau et al. 2014). Systematic differences in the magnitude scales of different observers amounting to a few 0\fm1 are normal. The last set of $BV$ photometry (2013--2018) is listed in \,Table\,I.1 of the Appendix.

  Only three mean quasi-periods could be derived (top of Fig.\,13), as reliable pulsations between 1995 and 2012 were lacking. As a consequence of the above, and the fact that after the 1975 eruption the star became heavily obscured (discussed in Sect.\,3.4.1.) no reliable correlation between mean quasi-periods and colour could be determined. On the contrary, the third mean quasi-period on the very right has become significantly shorter which is in agreement with the observed blueing trend until the last observations of 2018.

 The 1975 eruption named after Dean (1980) was only observed during the brightness rise after a deep minimum. The second eruption of HR\,5171A, including the deep minimum, was well observed, but only in the visual: the Otero minimum of 2000. The maximum just before the deep decline is the eruption maximum. Alas, multi-colour photometry and spectroscopy were lacking.

 The arrow at JD\,24\,47300, 1988 (see Fig.\,13), is a local minimum of the LTV during the $VBLUW$ photometric time series (Fig.\,3 upper panel), containing three pulsation cycles nos.\,15--17 (in Fig.\,1 of Paper\,I).
The deepest visual pulsation minimum is called the Jones-Williams minimum of 1995 (JD\,24\,49750, Fig.\,3 upper panel).
 The smoothed LTV in the $VBLUW$ system of Walraven is shown and discussed in\,Fig.\,J.1 of the Appendix.

\begin{figure}
 \resizebox{\hsize}{!}{\includegraphics[angle=-90] {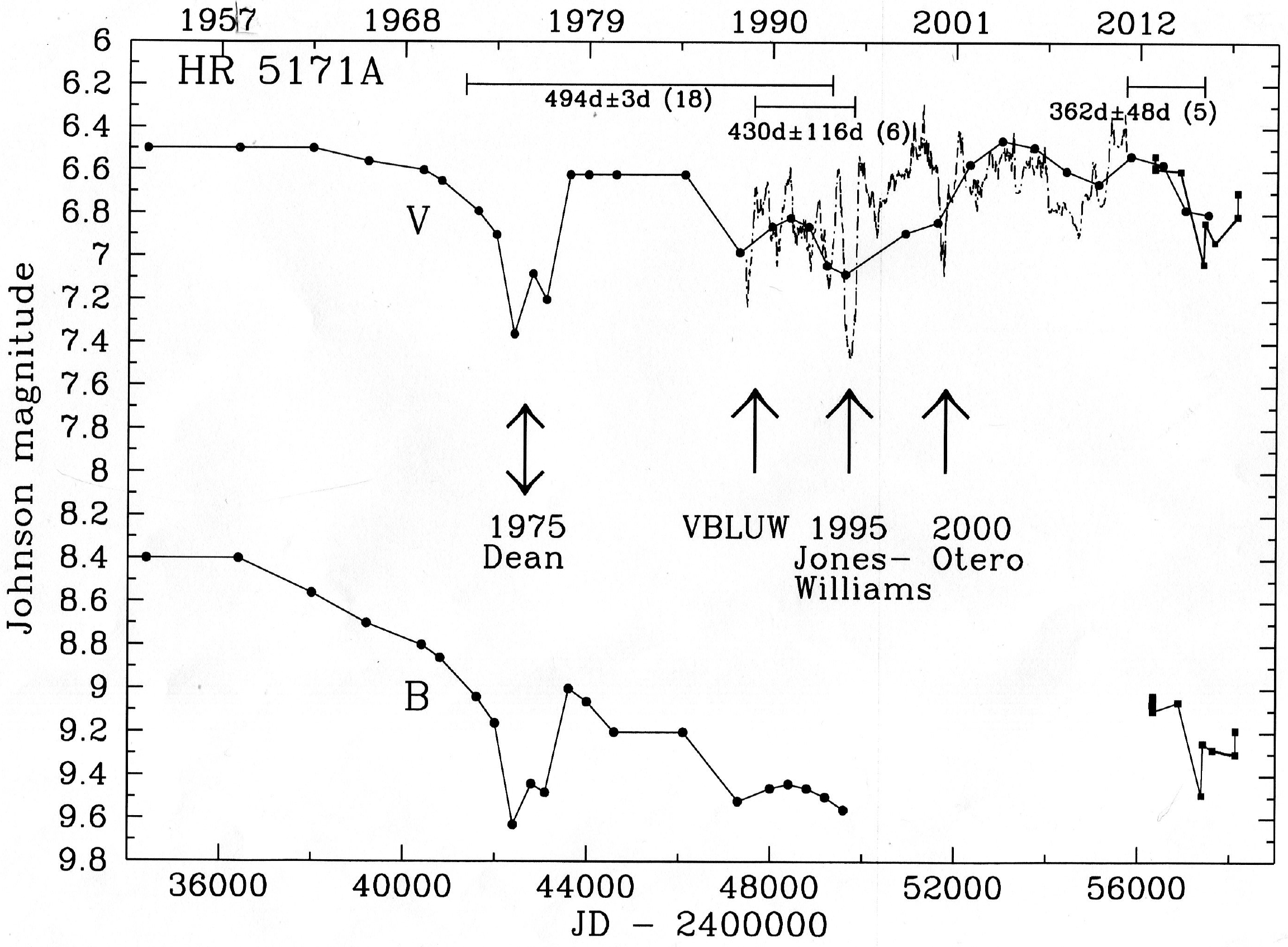}}
  \caption{Long-term variables for $V$ and $B$ of HR\,5171A 1953--2018. Time intervals (with mean quasi-periods and the sigma) are indicated by bars at the top. Bracketed numbers are the number of cycles. The mean quasi-period (494$^{\rm d}$) between 1970 and 1993 is based on a least-squares solution by van Leeuwen et al. (1998). See text for a discussion on the mean quasi-periods after $\sim$\,2000 (Sect.\,3.1.).
  The dashed light curve is based on the visual AAVSO data. The Otero-minimum is part of the massive 2000 eruption. Local brightness minima (of LTVs), or of individual pulsations are: the $VBLUW$ minimum (JD\,24\,47300), the Jones--Williams minimum (the deepest one at the time, JD\,24\,49750), and the Blown-minimum at JD\,24\,57330. The last two minima are identifiable in the  upper and lower panels of Fig.\,3, respectively. The squares on the very right are individual $BV$ observations made by GDS between 2013 and 2018 (Appendix\,Table\,I.1.); see Fig.\,15 for a detailed visual light curve in the same time interval. In this time interval ($\sim$\,JD\,24\,56755, 2014) the temperature was determined by Wittkovski et al. (2017a): 4290\,K\,$\pm$\,760\,K. }
  \label{Fig 13}
  \end{figure}

 \begin{figure}
  \resizebox{\hsize}{!}{\includegraphics[angle=-90]{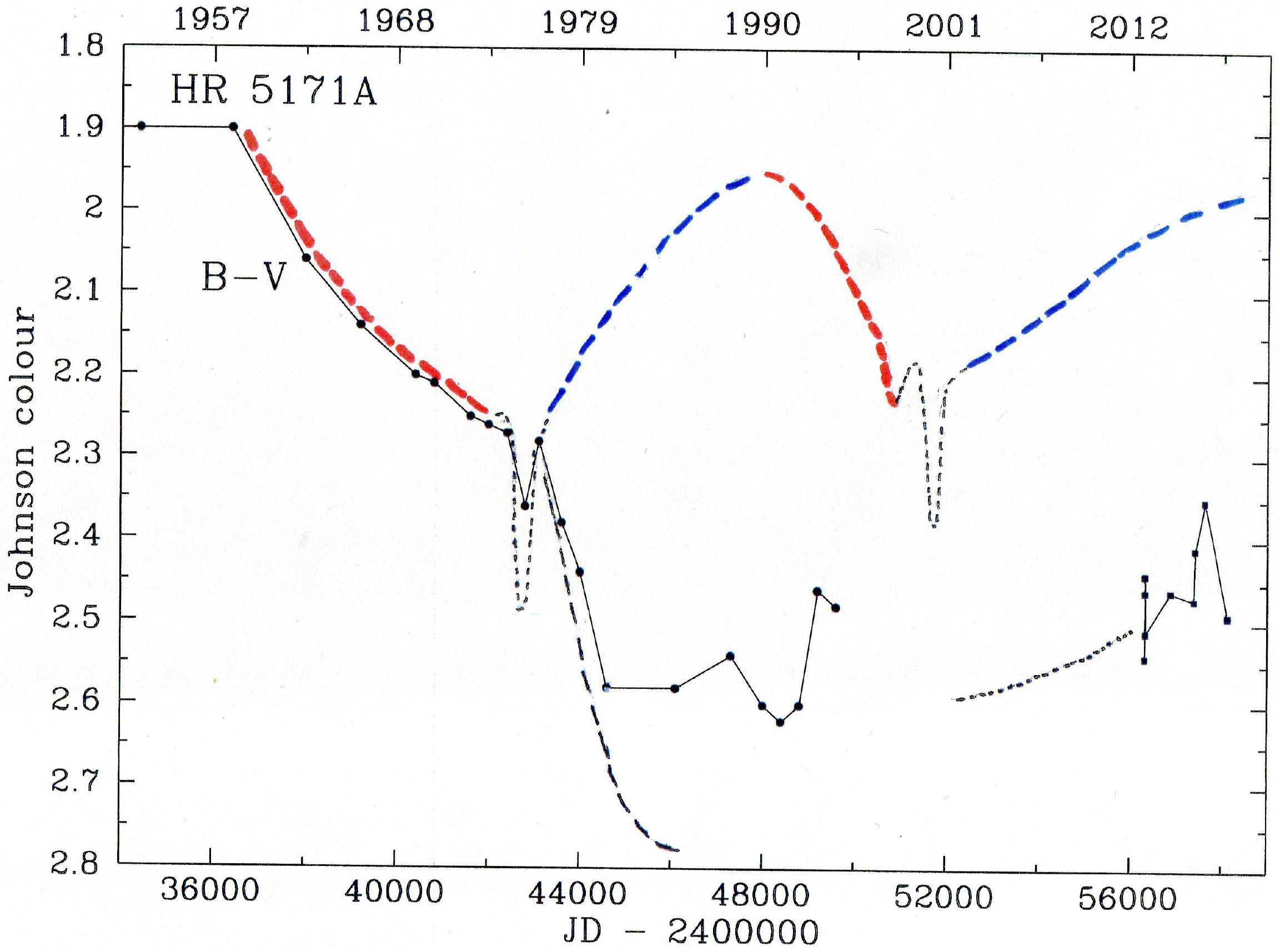}}
   \caption{Similar to Fig.\,13, but for the observed LTV of $(B-V)$ of HR\,5171A, 1953--2018: the black dots connected by black lines.  We note the long-lasting reddening episode between $\sim$\,JD\,24\,37000 (1960) and $\sim$\,JD\,24\,44500 (1981).
  All other curves are only schematic. In red and blue: the $(B-V)$ variations due to the RL and BL evolutions, respectively; black dotted peaks: the Dean 1975  and Otero 2000 eruptions; black dotted curve below and on the very right: the supposed blueing trend;
  black dashed curve running steeply downward: the supposed reddening by a shell ejection after the 1975 eruption. The result of the latter and the BL evolution above is the observed reddening of the LTV: the black dots connected by black lines between the 1975 eruption and $\sim$\,JD\,24\,46000.
  In the time interval between JD\,24\,50000 and JD\,24\,56000 (1994--2013) only visual observations were made. The small dataset (squares) on the very right (2013--2018) represents individual photoelectric observations by GDS (\,Table\,I.1 of the Appendix.); see  Fig.\,15 for the detailed visual light curve at the time.    }
    \label{Fig 14}
   \end{figure}

\subsection{Brightness and colour ranges of the red and blue 'evolutionary modes'.}
 Table\,J.1 of the Appendix presents a summary of the maximum ranges of the LTVs in brightness and colour index. These ranges serve a useful purpose, amongst others, giving a more precise determination of the actual evolutionary trends on the HR diagram in a later paper. Those of HD\,179821 are smallest. The largest ranges for the four objects occur in $U$ and $B$, that is, up to 1\fm2, and the smallest ones occur in $R$ and $I$. Those in $(U-B)$ are larger than in $(B-V)$.

 Another property is that the trends of the light variations in different pass bands often run in variable directions, and therefore in various `modes'. For example, $V$ and $B$ run in opposite directions and with different rates for a few years, then both decline in brightness with equal rates for another couple of years, after which $V$ declines much faster than the rise in $B$, and so on.

 We determined the evolutionary modes for HR\,8752, $\rho$\,Cas, and HD\,179821 and listed them chronologically in Table\,4, concentrating on those in $V$ and $B$ only, as $(B-V)$ is (apart from the reddening effect by the high-opacity layer) a temperature indicator. The red and blue evolutionary loops of HR\,8752 are shown in Fig.\,6 in Nieuwenhuijzen et al. 2012. The modes for HR\,5171A are collected in Table\,5, among which there are a few short-term modes.

 These tables list the type of modes with respect to the trends of $V$ and $B$ (brightness up, down, or constant), and of $(B-V)$ (to the red or to the blue, the latter in boldface), and the dates, JDs, and their durations.

 \begin{table*}
    \caption{Red and blue evolutionary modes for the LTVs of HR\,8752, $\rho$\,Cas, and HD\,179821 listed in chronological order. Second column: u\,=\,up\,=\, brightness increasing, d\,=\,down\,=\,brightness
  declining, if
  double, i.e. uu and dd, the increasing and declining, respectively, are stronger than in the other channel, c\,=\, constant. Third column:
  const.\,= \,constant. For clarity, the trends `blue' are in boldface. For HR\,8752 and $\rho$\,Cas, the transition from a red to a blue trend is caused by an eruption (1973, and 1986, 2000, 2013, respectively). Such a significant event for HD\,179821 should have happened, but was obviously missed. }
    \label{table:4}
    \centering
   \begin{tabular}{l l c c c c}
  \hline\hline
   Star  &  Mode &            Trend     & Dates & JD-      & Duration \\
         &                 & of $(B-V)$ &       & 24\,00000    & (d)     \\
  \hline
   HR\,8752    & $B_{\rm pg}$-d &         & 1899--1923 & 15000-23600 & 8600  \\
               & $B_{\rm pg}$-u &         & 1923--2001 & 23600-52000 & 28400 \\
               & $V$-uu B-u   & red  &  1942--1950 & 30500--33500 & 3000  \\
               & $V$-d $B$-d  & const & 1950--1957 & 33500--36000 & 2509 \\
               & $V$-d $B$-d  & red   & 1957--1976 & 36000--43000 & 7000  \\
               & $V$-d $B$-u  & {\bf blue} & 1976--2005 & 43000--53500 & 10500  \\
   \hline
   $\rho$\,Cas & $V$-u $B$-d  & red & 1965--1982 & 34000--45000 & 6000  \\
               & $V$-d $B$-dd & red & 1982--1986 & 45000--46500 & 1500  \\
               & $V$-u $B$-uu & {\bf blue} & 1986--1991 & 46500--48500 & 2000 \\
               & $V$-c $B$-d & red  & 1991--1998 & 48500--51000  & 2500 \\
               & $V$-d $B$-dd & red & 1998--2000 & 51000--51300  & 300  \\
               & $V$-c $B$-uu & {\bf blue} & 2001--2009 & 52000--55500  & 3500  \\
               & $V$-c $B$-dd & red  & 2009--2013 & 55000--56500 & 1500 \\
               & $V$-c $B$-u  & {\bf blue} & 2013--2015 & 56500--58100  & 1600  \\
  \hline
   HD\,179821  & $V$-d $B$-u & {\bf blue}  & 1990--1993 & 48000--49200  & 1200  \\
               & $V$-dd $B$-uu & {\bf blue} & 1993--2004 & 49200--53200 & 4000  \\
               & $V$-c $B$-d  & red &  2004--2008 & 53200--54500  & 1300  \\
  \hline
  \end{tabular}
  \end{table*}

\begin{table*}
 \caption{Similar to Table\,4, but now for HR\,5171A. The transitions between blue and red modes between 1973 and 1977, just before and after the 1975 eruption, are amazingly short: 200$^{\rm d}$--500$^{\rm d}$ compared to the other ones, and those of HR\,8752, $\rho$\,Cas, and HD\,179821 in Table\,4; see Sect.\,3.4.1. for a plausible cause. We note the gap between 1995 and 2013 which is caused by the lack of $BV$ photometry.
 The $BV$ data between 2013 and 2018 are listed in \,Table\,I.1 of the Appendix. The 2013--2018 mode is not yet finished, and therefore the duration is a lower limit   }
  \label{table:5}
  \centering
   \begin{tabular}{l c c c c }
   \hline\hline
   Mode & Trend      & Dates & JD-   & Duration  \\
        & of $(B-V)$ &       & 24\,00000   & (d)  \\
  \hline
  $V$-c $B$-c & const & 1953--1958 & 34400--36400 & 2000 \\
  $V$-c $B$-d & red   & 1958--1962 & 36400--37800 & 1400  \\
  $V$-d $B$-dd & red  & 1962--1965 & 37800--38800 & 1000  \\
  $V$-d $B$-dd & red  & 1965--1969 & 38800--40400 & 1600  \\
  $V$-d $B$-dd & red  & 1969--1973 & 40400--42000 & 1600  \\
 ($V$-dd $B$-d) & {\bf blue} & 1973--1975 & 42000--42500 & 500 \\
  $V$-uu $B$-u & red  & 1975--1976 & 42500--43000 & 500  \\
  $V$-u $B$-uu & {\bf blue} & 1976--1977 & 43000--43200 & 200  \\
  $V$-c $B$-d & red   & 1977--1981 & 43200--44800 & 1600  \\
  $V$-c $B$-c & const & 1981--1985 & 44800--46100 & 1300 \\
  $V$-d $B$-d & const & 1985--1988 & 46100--47300 & 1200  \\
  $V$-uu $B$-u & red  & 1988--1991 & 47300--48400 & 1100  \\
  $V$-dd $B$-d & {\bf blue} & 1991--1995 & 48400--50000 & 1600 \\

  $V$-dd $B$-d   & {\bf blue} & 2013--2018   & 56330--58140 & 1810 \\
\hline
\end{tabular}
\end{table*}

\subsection{Red and blue loop evolutions: their $(B-V)$ ranges and quasi-period trends.   }

Below we derive approximate evolutionary tracks, and an important quantity: the range in $(B-V)$ during the evolution, and thus the variation of T$_{\rm eff}$, assuming that it is almost independent of the atmospheric absorption (Sects.\,2.4.--2.6.). Additionally, we try to establish the true nature of HD\,179821 and discuss its probable isolated position in space.

The study of HR\,8752 by Nieuwenhuijzen et al. (2012) revealed that HR\,8752 evolved along a red loop evolution and after the 1973 eruption along a blue loop evolution (hereafter called RL and BL evolutions). Therefore, all red modes and all blue modes listed in Tables\,4 and 5 and preceding and following an eruption represent shorter pieces of one RL and one BL evolution, respectively.
This probably means that the evolution tracks on the HR diagram are not always straight, but show irregularities.
Below,  we firstly discuss these RL and BL evolutions of HR\,8752, $\rho$\,Cas, and HD\,179821.

{\bf --HR\,8752:} The construction of a RL from 1850 (best documented from 1895) until the 1973 eruption and subsequently the BL until 2005, are based on a mix of different observational techniques (Nieuwenhuijzen et al. 2014, their Table\,4 and Figs.\,5 to 24).
The $(B-V)$ ranges are 0\fm16 (1895--1963) and 0\fm64 (1976--2005). Between 1963 and 1976, no $BV$ photometry was available.

Observed and model timescales (by Meynet et al. 1994), that is, a few decades, are of the same order. Multi-colour photometry has been absent since 1993, but spectral observations were made until between 2000 and 2005 from which T$_{\rm eff}$ and log\,g from Kurucz's LTE models could be derived (Nieuwenhuijzen et al. 2012, their Table\,4 and Fig.\,2). During this BL evolution, temperatures and gravities fit the extrapolation line of the gradual rise of T$_{\rm eff}$ up to 8000\,K and log\,g from 1993 until 2005 very well.

Nieuwenhuijzen et al. (2012) claimed that HR\,8752 left the first instability region on the HR diagram, crossed the Yellow Void (Fig.\,1), and then was on its way to stability further to the blue, and that during further evolution it must still go through the second potential unstable region, where He starts to ionize. However, as mentioned before, the recent $BV$ photometry (by EJvB, Appendix\,B:2.1.) indicates that in 2017/2019 the $(B-V)_{0}$ of the star stopped blueing, but is still in the Yellow-Blue Void (Fig.\,1), and shows some variability in brightness and colour owing to its very-low-amplitude pulsations.
An additional analysis made by us of the AAVSO $V$ (Johnson) observations of HR\,8752 from 1993 until 2019 is presented in Appendix\,K:\,3.3.

{\bf --$\rho$\,Cas :} According to our collection of fragments of LTVs from numerous papers, $\rho$\,Cas has passed through three RL and two BL evolutions since the $BV$ photometry started in 1968; see the $(B-V)$ curve in Fig.\,12: a RL from 1968 to the 1986 eruption triggering a BL until about 1991, succeeded by a RL until the 2000 eruption, triggering the next BL until about 2008, then a RL until the end of the figure in 2018. Also, here the timescales are one to a few decades only.
The ranges in $(B-V)$ of two RL and two BL evolutions lie between 0\fm20 and 0\fm45, meaning that the temperature ranges can be derived after correction for reddening (Table\,1). They lie in the range of 5200\,K to 4500\,K (excluding the extreme temperatures during the eruption maxima and deep minima of about 7500\,K and 4000\,K, respectively).

 Based on the existence of such a `sequence of events', likely valid for all YHGs, we conclude that a third blue loop of $\rho$\,Cas must have taken place after the massive 1946 eruption (light curve in visual and photographic magnitudes by Gaposhkin 1949). Indeed, one to two decades later, in the 1960s and 1970s (in Figs.\,11 and 12 on the very left), the spectral type was F8, which is relatively blue (Humphreys 1978; de Jager et al.1988; Arellano Ferro \& Mendoza 1993). The LTV at the time $(B-V)_{0}$\,=\,0.60, which corresponds with T$_{\rm eff}$\,$\sim$\,6000\,K.

 According to Fig.\,12, the blue loop after the 2000 eruption reached roughly the same $(B-V)_{0}$ and consequently the same temperature as the eruption of 1946 until about 1968 (the latter date on the very left of Fig.\,12).
 Whether this was also the case for the blue loop after the 1986 eruption is uncertain, because of some inconsistency between the $(B-V)$ based on $BV$ photometry (dots in Fig.\,12) and the $(B-V)$ based on a transformation from the $VRI$ photometry (circles in Fig.\,12). It appears that an inconsistency exists between the trends of dots and circles which is disappointing, but usually the transformation yields more satisfactory results.

{\bf --HD\,179821 :} One glance at the light curves of this object was enough to recognize a BL evolution. It is staggering how closely the past history of HD\,179821 imitated that of HR\,8752, even if one compares the pulsation properties until 2018 made with the aid of the AAVSO Light-Curve Generator. The amplitudes and timescales of the quasi-periods are smaller and shorter (100$^{\rm d}$--150$^{\rm d}$), respectively, as far as the visual observations of HR\,8752 allow the comparison with the $UBV$ photometry by Arkhipova et al (2001, 2009), Le Coroller et al. (2003) of HD\,179821. The pulsations of both objects also share some photometric morphological properties.
Ikonnikova et al. (2018) reported that the quasi-periods of HD\,179821 became longer than 250\,d between 2010 and 2017, which can be expected when evolving along an RL evolution!

The BL of HD\,179821 started somewhere in the time interval 1925--1960, because at the time the star was very faint in B$_{\rm pg}$, and therefore very cool; see Fig.\,8.
The blue loop ended in 2004 (about JD\,24\,53000) when $(B-V)$ reddened again, marking a new RL evolution (Figs.\,9 and 10). The range of the colour $(B-V)$\,=\,0\fm22.

Given the characteristic `cyclic sequence of events' for HR\,8752 and HD\,179821, an eruption could have happened somewhere between 1925 and 1960, but like in the case of HR\,8752 (Sect.\,3.1.) stayed unnoticed.
The timescale is uncertain, but is approximately five decades, and therefore not in disagreement with theory
(the gaps in the time series of Fig.\,8 are in most cases much longer than the duration of eruptions: 400$^{\rm d}$--700$^{\rm d}$).

The bluest colour (corrected for IS extinction) of HD\,179821 in 2004 was $(B-V)_{0}$\,$\sim$\,0\fm55, its reddest colour in 1990, 0\fm77. Hence, the maximum photometric temperature is T(Phot)\,=\,5290\,K (dJN), while at the time the mean spectroscopic temperature according to Arkhipova et al. (2009) was T(Sp)= 6800\,K\,$\pm$\,50\,K (a difference of 1500\,K, see Sects.\,2.4.-- 2.6.). According to its position at the time in Fig.\,1, it stopped just before entering the Yellow-Blue Void and was heading back according to the available observations until 2009 (Fig.\,10).
With the Straizhys (1982) temperature calibration T(Phot) a comparable result was obtained by Arkhipova et al. (2009): 5400--6000\,K.
(As noted before new $UBV$ photometry by Ikonnikova et al. (2018), HD\,179821 continued its reddening trend until the end of the observations in 2017).

All the similarities above suggests that HD\,179821 is not a RSG, or a post-AGB star as is often claimed, but presumably a YHG at a very large distance (M$_{\rm bol}$\,$<$\,-8), supporting conclusions by Gledhill et al. 2002, Gledhill \& Takami 2001, Molster et al. 2002,  Arkhipova et al. (2009), and Oudmaijer et al. (2009). However, as a consequence, the large z-distance to the galactic plane creates kinematical problems, which can only be solved by a reliable trigonometric parallax (Arkhipova et al. 2009). Its large distance and its isolated position with respect to its place of birth (a young stellar cluster) is not unusual: see Smith \& Tombleson (2015: LBVs are `anti-social'). The isolated positions of many LBVs (the probable descendants of YHGs) are remarkable. Smith \& Tombleson believe that LBVs are flung out of their former binary orbit by a SN explosion. See also the related discussions by Humphreys et al. (2016) and Aghakhanloo et al. (2017). In the case of the suggested distance reduction for HR\,5171A, as discussed in Sect.\,2.9., its position also becomes isolated, as it will no longer be a member of the young stellar cluster Gum48d.

Our conclusions are that the observed $(B-V)$ ranges for the BL evolution lie roughly between 0\fm2 and 0\fm6, with T$_{\rm eff}$ roughly between 8000\,K and 4000\,K, roughly corrected for the too red $(B-V)$ values (Table\,A.1 of the Appendix).
Other novelties are that HD\,179821 is likely a YHG, and that all four YHGs seem to be isolated objects.

\subsection{HR\,5171A - a special YHG}

\subsubsection{The mysterious reddening episode 1960--1981.}

The most confusing property of HR\,5171A for a generation of researchers was the exceptional reddening episode by 0\fm7 in $(B-V)$ observed between about 1960 ($\sim$\,JD\,24\,36000) and 1981 ($\sim$\,JD\,24\,44000); see Figs.\,13 and 14. This happened before and after the 1975 eruption. After 1981 the reddening increase came to an end. Below we offer a plausible explanation with the aid of Fig.\,14.
(Its caption explains the meaning of the other curves: only schematically sketched).

Firstly, the long-lasting steep reddening episode starting around 1960 until 1981 cannot be caused by an increasing dust extinction, as the $B$ decline relative to that of $V$ happened much too quickly.
Secondly, because of the quick succession of three short-lasting modes: 500$^{\rm d}$, 200$^{\rm d}$ and 500$^{\rm d}$), a blue, a red, and a blue one, respectively, between 1973 and 1977 (Table\,5). Obviously, something exceptional happened at the time, disturbing the brightening phase after the 1975 eruption, as well as the two colour variations of the following pulsations in 1977 and 1978 (numbered 7 and 8 in Fig.\,1 in Paper\,I). As observations were scarce at the time, no reliable details can be offered.

However, as the three other YHGs highlight an alternating succession of a RL, an eruption, and a BL, with a timescale of a few decades only, this should also apply to HR\,5171A, but we are convinced that another event should have happened simultaneously.

We propose the following plausible scenario. The sketched alternating sequence of red and blue curves in Fig.\,14 represents the simplified and inevitable representation of the RL and BL evolutions, bounded to the two eruptions. We believe that the first observed reddening episode 1960--1974 was in fact a RL evolution, represented by the first red dashed curve, as it has a $(B-V)$ range of 0\fm35 (falling within the range of the other three YHGs, Sect.\,3.3.). Additionally, the $B$ decline relative to the decline of $V$ 1960--1974 in Fig.\,13 is quite unlike an increasing extinction by more dust.

The reddening in the second episode 1976--1981 started right after the 1975 eruption; it increased even faster than the one between 1960--1974, and stopped in 1981. This is quite awkward. It is plausible that the latter reddening episode was mainly due to a massive shell ejection during the 1975 eruption, schematically represented by the black dashed curve in Fig.\,14. The $B$ absorption was obviously much higher than in $V$. This event happened simultaneously with the first BL evolution, starting right after the 1975 eruption. The result of both events produced the observed LTV represented by the black dots connected by straight lines.

A shell event happening right after the 1975 eruption is likely considering the existence of a massive optically thick extended molecular and dust envelope around HR\,5171A; see Sects.\,2.9 and 3.4.2. Obviously, such events happened in the past as well.

This first BL evolution was succeeded by a new (second) RL evolution bounded by the impending eruption in 2000. Hence, the observed $(B-V)$ in Fig.\,14 became bluer as $V$ declined more than $B$ between 1990 and 1995; see Fig.\,13. We note that this probably explains the relatively large depth of the Jones--Williams pulsation minimum at JD\,24\,9750 (1995); see Figs.\,3, and 13.

The continuation of the dashed black curve to the right at the bottom of Fig.\,14 roughly indicates the presumable continuation of the intrinsic reddening by the shell, and its subsequent decrease. After the 2000 eruption, a second BL evolution then developed which is responsible for the blueing trend of $(B-V)$ after 2001, and was still going on early 2019 (Sect.\,3.4.3.).
In conclusion, here we offer a new and very plausible explanation for the persistent reddening of HR\,5171A between 1960 and 1981.

\begin{figure}
 \resizebox{\hsize}{!}{\includegraphics[angle=-90]{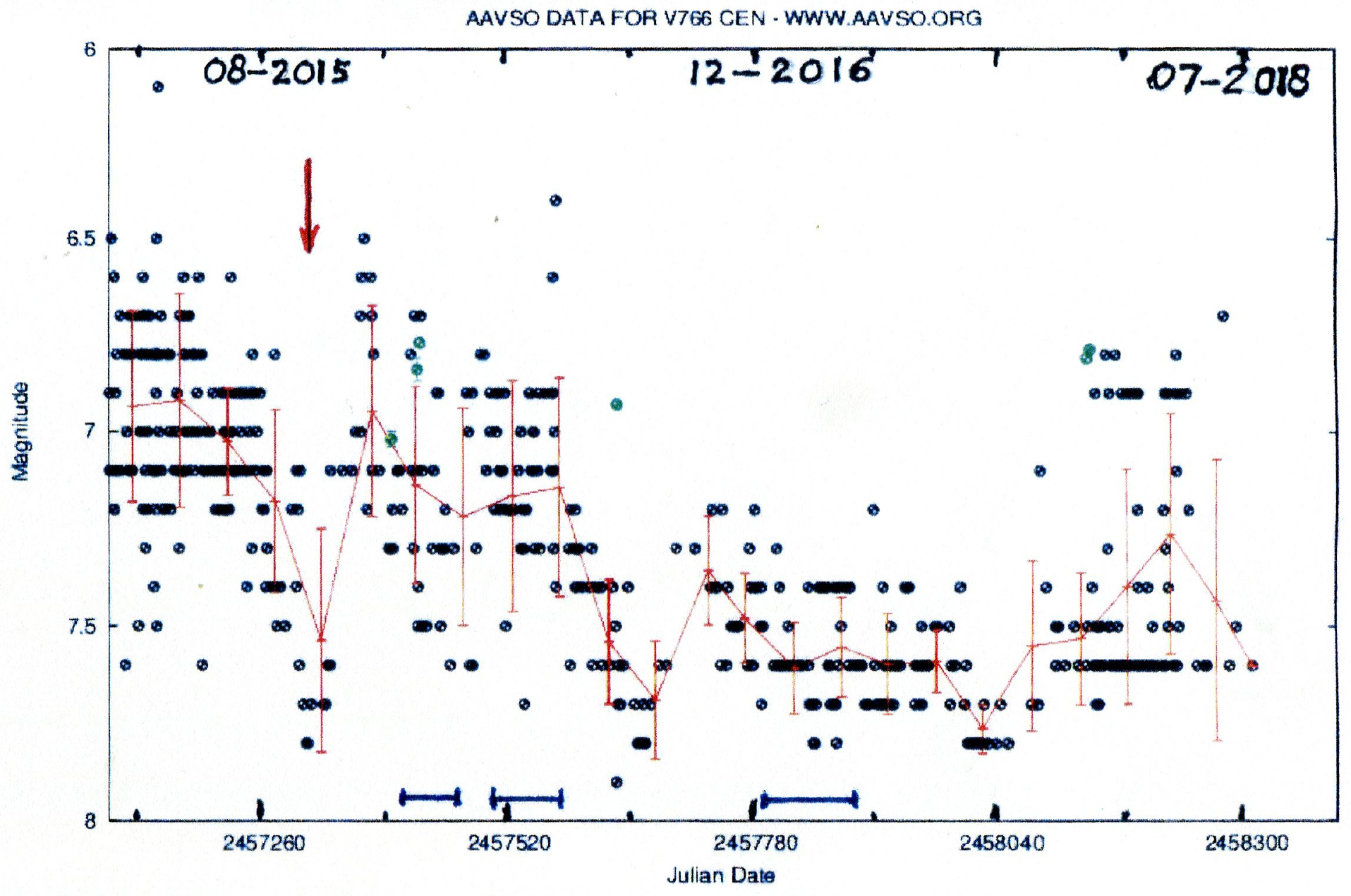}}
  \caption{Visual AAVSO light curve 2011--2018 (red lines based on 50$^{\rm }$ d averages) of HR\,5171A showing the last normal pulsation with the Blown-minimum (red arrow at $\sim$\,JD\,24\,57330, early October 2015, cf.  second
panel of Fig.\,3), during an observed portion of the BL evolution showing sudden decreasing quasi-periods and declining light amplitudes. The behaviour after JD\,24\,57650 is similar to those exhibited by HR\,8752 and HD\,179821 during their BLs. The green dots are the photo-electric $V$ observations from Table\,I.1, which are brighter than the visual estimates by 0\fm2--0\fm4 as expected. The visual ones should be shifted upwards, but for clarity this was not done. See text for further explanation.  }
   \label{Fig.15}
  \end{figure}

\subsubsection{The 1977 excess of Balmer continuum radiation of HR\,5171A and the near-infrared $JHKL$ connection. }

Firstly, we relate the Balmer continuum excess of HR\,5171A observed in 1977 to its 1975 eruption episode and shell event with the aid of the $VBLUW$ photometric system. Subsequently, the effects of RL and BL evolutions and the 1975 eruption on the $JHKL$ light curves are discussed.

One can only speculate as to any relation between the matter blown into space after the atmospheric explosion of 1975 and the anomalous high UV excess happening during the development of the very-high-amplitude pulsation of 1977 immediately following the eruption (labelled no.\,7 in Fig.\,1 of Paper\,I). The Balmer continuum radiation excess amounted to about 1$^{\rm m}$ (see Fig.\,1 in van Genderen et al.\,2015). No such excess was observed in 1971, 1973, or between 1980 and 1991 when also $VBLUW$ photometry was made.
A glance at the $B_{\rm pg}$ light curve of the Harvard DASCH scanning project (not shown, only discussed in Sect.\,3.1.) indicates that such deep brightness declines could have happened in the past as well. (Maximum light was not observed during the 1975 eruptive episode, nor was the deep brightness decline to a deep minimum; only the steep brightness rise was seen thereafter: see Fig.\,13.).

Nevertheless, after 1981 the observed $(B-V)$ curve until the end of our time series in 2019 remained overly red by approximately 0\fm6  compared to simplified RL and BL evolutions, of which the brightness level in Fig.\,14 should be more or less correct. A causal connection with the 1975\,\,eruption and the effects of its abundant mass-loss is plausible. It is not surprising that HR\,5171A is imbedded in an extended dust and molecular envelope (Chesneau et al. 2014; Wittkovski et al. 2017a, and Sect.\,2.9). We note that an overall long-term variation of the Balmer continuum radiation exists (Appendix\,J:3.4.2. and Fig.\,J.1.).

The NIR $JHKL$ photometric light and colour curves, obtained at the SAAO and presented and discussed by Chesneau et al. (2014) in their Figs.\,4 and 7, clearly show the same BL and RL evolutions after the 1975 eruption until 2012, just like the ones sketched in our Fig.\,14. The first BL evolution reached maximum brightness at 1990, just like in our sketch of Fig.\,14 at the bluest $(B-V)$. The 2000 eruption was reached after a RL with a brightness decline of 0\fm6 in $L$ and $K$, but less in $H$ and $J$.
The NIR 1975 eruption minimum was at least 0\fm3 fainter on average than that of the 2000 eruption. It is a pity that at the time of the latter, no $B$ magnitudes were made, meaning that the $(B-V)$ trend until 2012 is unknown (Fig.\,14).
The trend of the colour indices was discussed by Chesneau et al. (2014: Sect.\,4.1.).

In conclusion: HR\,5171A was likely subject to a massive shell event after the 1975 eruption.  Blue- and red-loop evolutions appear to also be detectable in the NIR photometry.

\subsubsection{Current BL evolution of HR\,5171A.}

In the autumn of 2015, our attention was aroused by the frequent reports of one of the present authors (EB) on the behaviour of HR\,5171A based on his monitoring of the visual brightness and supported by many AAVSO colleagues. They are Alexandre Amorim, Brazil; Giorgio Di Scala, Australia; Hiroshi Matsuyama, Japan; Andrew Pearce, Australia; Peter Reinhard, Austria; Eric Blown, New Zealand; Eduardo Goncalves, Brazil; Alan Plummer, Australia; Antonio Padilla Filho, Brazil; Peter Williams, Australia: the visual brightness decline of a pulsation to a minimum in the autumn of 2015 seemed to happen relatively quickly (called the Blown-minimum at JD\,24\,57300, October 2015); see red arrow.
The pulsation amplitudes became suddenly smaller after 250 d (JD\,24\,57520), and the quasi-periods shorter until the end of our dataset at JD\,24\,58200 (March 2018), a time interval of about 2\,yr. At the same time $(B-V)$, based on $BV$ photometry shown in \,Table\,I.1 of the Appendix, became bluer -- a similar behaviour to that of HR\,8752 and $\rho$\,Cas.

It appeared that from the end of December 2015 to 2018, the light curve plotted by the AAVSO Light Curve Generator (JD\,24\,57100--JD\,24\,58300, March 2015--July 2018), shown in Fig.\,15, revealed a most astonishing pattern: a mean decrease of the visual brightness (see Fig.\,3 second panel of which Fig\,15 represents the continuation), smaller light amplitudes, and much shorter quasi-periods within a time span of only $\sim$\,250\,d. This happened between the maximum at JD\,24\,57400, right after the Blown-minimum, until JD\,24\,57650. During this pulsation, including the Blown-minimum and the maximum afterwards, some spectra were taken. An echelle spectrum was obtained, including H$\alpha$ by E. Budding (ed.budding@gmail.com), and a high resolution spectrum was obtained by Dr. N.I. Morrell (nmorrell@lco.cl), but there were no abnormal features; the spectra looked normal (private comm. 2015).

The three bars at the bottom of Fig.\,15 mark the epochs for the three $H$-band images of HR\,5171A made by Wittkovski et al. (2017c). These images show a variable circumference, we presume not of the photosphere, but of its NIR $H$ band-radiating layer (although we expect the photosphere to be disfigured as well).

In conclusion, thanks to the continuous visual observations, we detected a surprising feature:
during the BL evolution in 2016, the pulsation amplitudes of HR\,5171A suddenly appeared much smaller (and probably also with shorter periods). This phenomenon has not been observed before.

\begin{figure}
   \resizebox{\hsize}{!}{\includegraphics{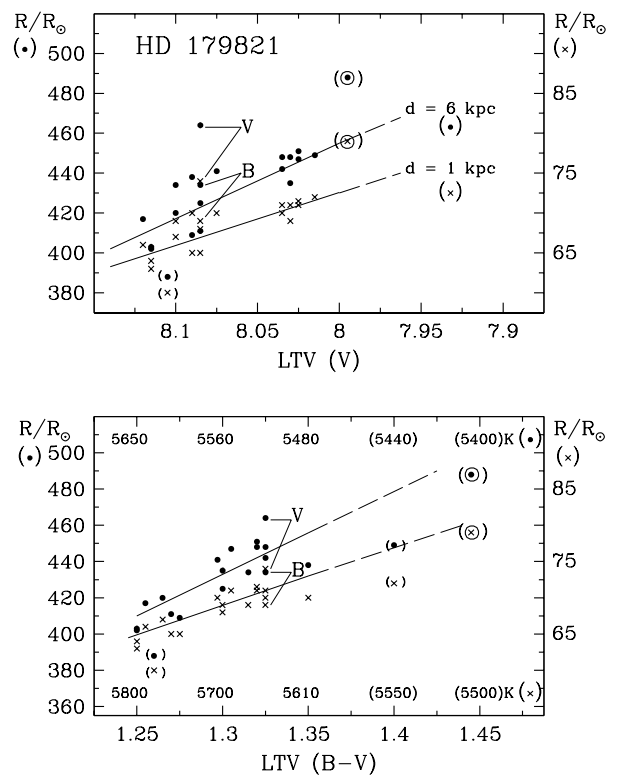}}
   \caption{Demonstration of how T$_{\rm eff}$(Phot) and R/R$_{\rm \odot}$ of the light maxima of HD\,179821 depend on the brightness and colour of the blue loop evolution. The stellar parameters of photometric temperature and radius variations were calculated for two possible distances 1\,kpc (x signs) and 6\,kpc (dots) using the dJN method. Brackets indicate uncertain results by a lack of data points in the light curves, encircled symbols represent the pulsation maximum at
   JD\,24\,48550 of which $V$ is very bright, but the colour indices $(B-V)$ and especially the $(U-B)$ extremely red. The symbols labelled $V$ and $B$ refer to the maximum of a very peculiar pulsation at $\sim$\,JD\,24\,49300, of which the BL evolution in $V$ declined significantly sooner than in $B$. As most light maxima belong to a blue loop evolution (between JD\, 24\,48000 and JD\,24\,53000), Figs.\,9 and 10, the photometric temperatures (lower panel) increase in concert with decreasing radii, and a declining visual brightness (upper panel).}
 \label{Fig. 16}
  \end{figure}

\subsection{ Radius variation and contraction rate of HD\,179821 during the BL evolution.}

Below we explore the effects of a BL evolution on the radius, the T(Phot), and the contraction rate, which has never been done before.

As we are dealing with evolutionary tracks on the HR diagram, we take it for granted that the stellar property variations keep pace with the temperature. Hence, one expects that radial velocity curves also show LTVs. Indeed, radial velocity monitoring campaigns of $\rho$\,Cas (monitored by Lobel et al. 2003, Figs.\,1 and 4) reveal LTVs as well, and as far as we can judge these are more or less in concert with those of the visual brightness.

As long-term radial velocities for HD\,179821 are lacking, we explored the effects of the BL evolution on its radii and photometric temperatures. We note that a calculated T(Phot) is always lower than a T(Sp). For example, an overly red $(B-V)$ is responsible for an overly low temperature, and according to the formula log\,R/R$_{\rm \odot}$\,=\,8.48--0.2\,M$_{\rm bol}$--2\,log\,T$_{\rm eff}$, the radius becomes larger for an almost constant M$_{\rm bol}$. Therefore, the results should not be taken too literally, as we are only interested in the general effects of a BL evolution on the stellar properties.

The illuminating result is presented in Fig.\,16. The distance of HD\,179821 is a serious matter of debate: between $\sim$\,1\,kpc and $\sim$\,6\,kpc (e.g. Surendinanath et al. 2002; Arkhipova et al. 2001, 2009; Le Coroller et al. 2003; Nordhaus et al. 2008; Patel et al. 2008; Oudmaijer et al. 2009; Ferguson \& Ueta 2010; Sahin et al. 2016), M$_{\rm bol}$\,$\sim$\,-4.3 and -8.2, respectively, but according to our conclusion in Sect.\,3.3., a large distance is favoured.

We calculated the trend of R/R$_{\rm \odot}$ as a function of the $V$ and $(B-V)$ of the LTVs (here
only a blue loop track, of which the photometry is depicted in Figs.\,9 and
10, respectively) for most of the pulsation maxima of HD\,179821, and for both distances. For the dJN method we then need to differentiate between an Ib--II supergiant and a hypergiant between the Ia$^{+}$ and Ia luminosity classes, for the short and long distances, respectively.
Averages of the s-parameter for the two pairs of luminosity classes were used to derive T$_{\rm eff}$ and BC (Appendix\,F:2.7.). The extinction and reddening used are from Table\,1.
On the left (dots) the trend of the radii for 6\,kpc on the right (crosses) for 1\,kpc.
The scatter around the mean lines is largely intrinsic. This can be explained as follows. About 30\% of the light ($V$) and colour ($(B-V)$) curves of HD\,179821 do not run in phase. As a consequence, at a fixed $V$ brightness, a great variety of $(B-V)$ colours exist, and therefore also a great variety of temperatures. According to the formula above, this causes a scatter as well in the computed radii, while the effect of the variation of M$_{\rm bol}$ is small.

The radii at maximum light vary from 400 to 450\,R$_{\rm \odot}$ and 60 to 80\,R$_{\rm \odot}$ for 6 and 1 \,kpc, respectively.
 The trends in Fig.\,16 from the right to the left indicate a decrease of the radii for the distances d\,=\,6\,kpc and 1\,kpc, by about 60\,R$_{\rm \odot}$ (15\%) and
 11\,R$_{\rm \odot}$ (15\%) by an LTV visual brightness decrease of 0\fm11 in $V$, and an LTV blueing of 0\fm15 in $(B-V)$, respectively.

We derived the contraction rate during the BL evolution of HD\,179821 (roughly between JD\,24\,49000 and JD\,24\,53200) for a distance of 6\,kpc: the result is $\sim$\,2.0\,R$_{\rm \odot}$\,y$^{-1}$.

\section{Theoretical and speculative explanations for two observational highlights.}

Below we discuss two speculative subjects. Firstly, the one raised in Sect.\,2.3.
We  believe that all convective stars, such as YHGs, are subject to a random process that may be identified as the convection dynamics acting on the periodic pressure waves by the $\kappa$-mechanism of the He-ionization zone (Fadeyev 2011). This principle has been used by Icke et al. (1992) for their numerical study of the instability of post-AGB stars, producing model light curves with some similar characteristics to those of YHGs.

It should be noted that Wisse (1979) and Icke et al. (1992) did not include the LTVs (representing the BL and RL evolutions) in their calculations, on which the much shorter quasi-periods are superimposed. This means that the calculations should have been done with one more variable. Based on a comparison of the main characteristics of the observed time series of the four YHGs, and those of the calculated time series without LTVs, we conclude that LTVs have obviously little effect: the observed pulsations also show a weakly chaotic character just like the models.

Now we discuss observations detailed in Sects.\,2.4.--2.6. and in Appendix\,D:\,2.6., namely indications for the presence of a variable high-opacity layer in the atmospheres of YHGs, responsible for a wavelength-dependent absorption.
Deviating photometric and calculated stellar properties emerged from analyses of individual light and colour curves, for example by means of the $V$/$(B-V)$ and Ampl\,$V$/Ampl\,$(B-V)$ diagrams. As spectral temperatures are independent of continuum absorption, they always appear to be higher than photometric ones. The differences decrease from about 3000\,K to a few 100\,K from high (about 8000\,K) to low (about 4500\,K) temperatures, respectively.

As a result of the abnormally high amplitude ratios Ampl\,$V$/Ampl\,$(B-V)$, the calculated temperatures for HR\,8752, for example, are about 200\,K too low (compared to the results of Nieuwenhuijzen et al. 2012: their Fig.\,2 based on MK spectral types and LTE effective temperatures), and the calculated stellar radii in the maximum are largest: 946\,R/R$_{\rm \odot}$ and in the minimum smallest: 866\,R/R$_{\rm \odot}$, respectively. The dimensions on the other hand should be the other way around (estimated relative and absolute errors are about 100\,R/R$_{\rm \odot}$, compared to the results of Nieuwenhuijzen et al. (2012) in their Fig.\,20).
The photometric observations on which our calculations are based, are from Walker (1983); Ferland priv. comm. to Lambert \& Luck (1978); Moffett \& Barnes (1979).

A discussion on variable transparency sources in stellar atmospheres is given by Lobel (1997, Ch.1.).
For example, it is known that the opacity depends on the density of partially ionized hydrogen gas. The quasi-periodic pulsations of YHGs expand and compress their very extended atmospheres causing variable opacity and absorption of radiation that yield temporal changes of the stellar T$_{\rm eff}$ and mass-loss rate. During the pulsation-compressions the increased surface gravity raises the gas density of extended layers in the optically thin portions of the outer atmosphere (i.e. $\tau_{\rm R}$<$2/3$ where the local radiation temperature T$_{\rm rad}$ is not in equilibrium with T$_{\rm gas}$), consequently altering the selective absorption of incoming radiation and producing significant optical depth effects.
Lobel (2001) discussed the possibility that the T$_{\rm eff}$ variations with pulsations can alter the temperature of the incoming radiation field (with T$_{\rm rad}$ variations proportional to the T$_{\rm eff}$ changes) inside these atmospheric layers, further shifting the local (LTE) thermal H ionization balance due to (non-LTE) photo-ionization. The deviations from an average T$_{\rm rad}$ occur because of variations in the local opacity sources at different atmospheric heights. The modified partial H ionization alters the layer opacity. But it also changes the overall compressibility (or bulk modulus) through $\Gamma_{1}$ and hence its local dynamic stability in the stellar atmosphere that may cause the unreliability of $(B-V)$ to derive an accurate temperature.

 \section{Conclusions.}

The most important novelties to come from the present study can be summarised as follows.

1. Spectroscopic temperatures (T(Sp)) are always higher than photometric temperatures (T(phot)) independent of the temperature scale used. Differences (a few thousand K near 8000\,K) decrease towards lower temperatures (a few hundred K near 4200\,K), and must be due to the presence of a temperature-dependent high-opacity layer in the atmosphere. After all, T(Sp) is relatively insensitive to continuum absorption, contrary to T(Phot). As a result, $B$ and $V$ magnitudes are inconsistent with respect to each other; $B$ has suffered from more absorption than $V$.
Obviously, the higher the temperature, the higher the opacity of that atmospheric layer. It goes without saying that this was caused by increased mass loss and subsequently by a higher gas density of that layer rendering a too low T(Phot) (Sects.\,2.4.--2.6.).

2. The analysis of the LTVs of $\rho$\,Cas, HD\,179821, and HR\,5171A, and confirming the LTV of HR\,8752 by Nieuwenhuijzen et al. 2012, revealed that they mainly consist of RL and BL evolutions. These evolutions are responsible for the alternating T$_{\rm eff}$ variations of 1000\,K--4000\,K. The corresponding observed $(B-V)$ ranges are 0\fm2--0\fm6 (Sects.\,3.2. and 3.4.1.).

3. The above suggests that HD\,179821 is likely a YHG, not a RSG, nor a post-AGB star. If correct, the star should be located at a large distance (probably larger than 5\,kpc), and far above the galactic plane, creating a kinematical problem (Sect.\,3.3.) as already mentioned by Arkhipova et al. (2009).

4. In light of the results described in (2) above, a schematic sequence of RL and BL evolutions could be defined for HR\,5171A in Fig.\,14, separated by the 1975 and 2000 eruptions. The most plausible explanation for the persistent reddening episode in 1960--1981 of HR\,5171A is that it consists of an RL evolution between 1960 and the 1975 eruption. Thereafter, a severe absorption episode should have happened right after this eruption by a massive shell ejection, simultaneously with a BL evolution (Sect.\,3.4.1.).

5.  Quasi-periods, pulsation amplitudes, and stellar radii decrease and increase during BL and RL evolutions, respectively (Sects.\,2.5., 3.1., 3.3, 3.4.3., 3.5., Figs.\,11 and 13 and Appendix\,K:3.3.).

6. Based on light-curve markers, a new but weak eruption of $\rho$\,Cas was identified which happened in 2013, and which was independently spectroscopically confirmed by Aret et al. (2016; Sect. 2.2. in the present paper).

7. The light variations of HR\,5171A of the order of a year are due to pulsations alone. Any significant contribution to the light variation of a distorted primary by a possible much smaller contact companion is out of the question (Sect.\,2.3.; Chesneau et al. 2014, their Fig.\,8). However, we cannot exclude an unobserved companion.

8. The photometric observations reveal that each pulsation is unique, any predictability on the morphology and duration of the next pulsation is impossible. These facts are in accordance with the light-curve models of Icke et al. (1992) and Wisse (1979) for evolved long-period variables (Sect.\,2.3.). Based on the model light curves of  Icke et al. (1992), we characterize the pulsations of the four YHGs as `weakly chaotic' rather than `chaotic'. For both types of model light curves, the incessant sequences of photometric pulsations appear coherent, which agrees with the observations (Figs.\,2 and 3).

9. The membership of HR\,5171A to the stellar cluster Gum48d at a distance of 3.6\,kpc is doubtful. Its distance is probably of the order of 1.5\,kpc\,$\pm$\,0.5\,kpc (Sect.\,2.9.). If this supposition is correct, it would prove that HR\,5171A is not amongst the biggest stars known. For $\rho$\,Cas, a shorter distance is preferred as well: 2.5\,kpc\,$\pm$\,0.5\,kpc, instead of the accepted one of 3.1\,kpc\,$\pm$\,0.5\,kpc, (Sect.\,2.9.).

10. We would like to finish this paper with the motto of the Royal Astronomical Society, London, 1820:
`Quicquid nitet notandum', or `Whatever shines should be observed'.
See Appendix\,L: Message to observers.

\begin{acknowledgements}
A.L. acknowledges support in part by the Belgian Federal Science Policy Office under contract No.\,BR/143/A2/BRASS and funding received from the European Union's Framework Programme for Research and Innovation Horizon 2020 (2014--2020) under the Marie Sklodowska-Curie grant Agreement No.\,823734.
G.W.H. thanks Lou Boyd for his support of Fairborn Observatory and acknowledges long-term support from Tennessee State University and the State of Tennessee through its Centers of Excellence Program.
We acknowledge with thanks all the variable star observers worldwide from the AAVSO, especially S. Otero, VSX Team, for some invaluable comments on this paper and offering his historical data of HR\,5171A which were used in this paper, and which will be of great benefit for future research.
We like to thank Mr. Roland Timmerman for making Fig.\,1.
An anonymous referee is gratefully acknowledged for the invaluable comments and suggestions, which improved the presentation of this paper.

We would like to thank the following colleagues for their stimulating and fruitful correspondence since 2013, on various subjects addressed in this paper: Prof.dr. R.M. Humphreys, Dr. R.D. Oudmaijer, Dr. F. van Leeuwen (Hipparcos), Dr. B. Mason (Washington Double Star Cat.), Dr. J. Grindlay (Harvard DASCH scanner), Prof. dr. V. Icke, Dr. J. Lub, Prof.dr. J.W. Pel, Dr. D. Terrell, Prof.dr. A. Witt (Blue Luminescence of PAH molecules), Mr. E. Budding and Dr. N. I. Morrell (spectrograms of HR\,5171A), Dr. C. Sterken, Dr. A. Meilland and Dr. E. Chapellier. This research made use of the excellent near-infrared (JHKL) data obtained at the SAAO, and discussed by Chesneau et al. (2014).

  \end{acknowledgements}

   \listofobjects

   \begin{appendix}

   \section{1. Data summary.}
    See Table\,A.1.
\begin{table*}
  \caption{Data for the four YHGs used in Fig.\,1.
The distances used are for HR\,8752: 1.370\,kpc\,$\pm$\,0.480\,kpc (Nieuwenhuijzen et al. 2012: based on the Hipparcos parallax: van Leeuwen 2007), $\rho$\,Cas: 2.5\,kpc\,$\pm$\,0.5\,kpc (Sect.2.9.), HR\,5171A: 1.5\,kpc\,$\pm$\,0.5\,kpc (Sect.\,2.9.), and HD\,179821: 6\,kpc (Sect.3.3.). We preferred to use spectroscopic temperatures if available: the four of HR\,8752, and the second of HD\,179821. If spectroscopic temperatures were not available, we used temperatures based on the less reliable $(B-V)$ (Sects.\,2.4.--2.6.).}

   \label{table:A.1.}
    \centering
    \begin{tabular}{l c c c c }
    \hline\hline
    Star     & Time & log\,T$_{\rm eff}$ & log\,L/L$_{\rm \odot}$ & $(B-V)$  \\
    RL or BL & interval & range          & range                & range   \\
    \hline
     HR\,8752 &     &                    &                      &    \\
     RL      & 1895--1963  & 3.72--3.70  & 5.25--5.38           & 0\fm16  \\
     BL      & 1976--2005  & 3.70--3,90  & 5.60--5.36           & 0\fm64  \\
     \hline
     $\rho$\,Cas  &     &                &                      &         \\
     RL      & 1968--1998  &  3.72--3.66  & 5.48--5.54          & 0\fm40  \\
     \hline
     HR\,5171A &         &                &                     &         \\
     RL      & 1958--1974  &  3.70--3.65  & 5.40--5.30          & 0\fm35  \\
     \hline
     HD\,179821 &           &             &                     &         \\
     BL      &  1990--2004  & 3.69--3.83  &  5.20--5.10         & 0\fm22   \\
     \hline
     \end{tabular}
     \end{table*}

   \section{2.1. Databases. }
 Excellent and homogeneous time-series exist for $\rho$\,Cas initiated by Henry (1995, 1999; see footnote\,No.1): 1986--2001 ($VRI$) and 2003--2018 ($BV$), with a time resolution down to about 1$^{\rm }$ d and often with an accuracy of 0\fm02--0\fm01.

The longest recorded time-series also including eruptions are those made for $\rho$\,Cas by visual observers since the 1960s, summarised by  Beardsley (1961) for example. Others are Bailey (1978) and the AAVSO observers. A great amount of $(U)BV$ photometry was performed until the early 1990s for example by Brodskaya (1966), Fernie et al. (1972), Landolt (1968, 1973), Percy \& Welch (1981), Leiker \& Hoff (1987, 1990), Leiker et al. (1988, 1989, 1991), Halbedel (1985, 1986, 1988a, 1991b, 1993a), Henry (1995, see paragraph above), Percy et al. (1993, 2000), Zsoldos \& Percy (1991), and by the Hipparcos satellite 1989-1993 (ESA 1997). A number of $BV$ observations were obtained between 2016 and 2018 by one of the current authors: (EJvB) and plotted in Fig.\,2. The observations were made with a 10" f/6.3, Meade LX200 telescope. See for the observations: https://www.aavso.org/data-access, and for the most recent ones: in window `Pick a star' on www.aavso.org.

HR\,8752 was rather intensively monitored during numerous observing campaigns in the eighties until 1993 in $BV$, but less in $U$ ($UBV$): Arellano Ferro (1985), Halbedel (1985, 1986, 1988a, 1991a, 1993b), Mantegazza et al. (1988), Moffett \& Barnes (1979), Parsons \& Montemayor (1982), Percy \& Welch (1981), Walker (1983), Zsoldos \& Olah (1985), and by the Hipparcos satellite 1989--1993 (ESA 1997). Lastly, a number of $BV$ observations were made between 2016 and 2019 by one of the present authors (EJvB, see above).

Splendid time-series for HD\,179821 were obtained by: Arkhipova et al. (2001, 2009, $UBV$) between 1990 and 2008, who also retrieved photographic magnitudes for the time interval 1899-1989, by Le Coroller et al. (2003: $UBVRI$) and by Hrivnak et al. (2001: $V$), partly simultaneously with the previous groups.

HR\,5171A was monitored in various multi-colour photometric systems: 1953-2016, collected from literature in Paper\,I and by Chesneau et al. 2014, by the Long-Term Photometry of Variables project at La Silla (LTPV) organised by Sterken 1983, see Manfroid et al. 1991, Sterken et al. 1993; by the Hipparcos satellite, ESA 1997, see van Leeuwen et al. 1998; and by one of the present authors (GDS: Appendix Table I:1.). A large body of splendid $JHKL$ data was obtained at the SAAO from 1975 until 2013, published and discussed by Chesneau et al. (2014).

The AAVSO and ASAS observers appeared to be very productive for the Big Three ($\rho$\,Cas, HR\,8752 and HR\,5171A), especially in the last three decades; very valuable sequences of pulsations were recorded.

 \section{2.2. Quasi-period search.}
Particularly in the 1980s and 1990s, the type of the quasi-periodic pulsations, radial or non-radial, was heavily debated. Models indicated that they should be non-radial (Maeder 1980; de Jager 1993, 1998; Fadeyev 2011; Lobel 1997: Ch.2; Lobel et al. 1994. It is believed that they are probably due to gravity waves (de Jager et al. 1991)

The many attempts to find a consistent quasi-period, and to decipher whether they are radial or non-radial pulsations, has not lead to cogent results (e.g. Arellano Ferro 1985; Halbedel 1991b; Zsoldos \& Percy 1991; Percy et al. 2000; Arkhipova et al. 2001).
Le Coroller et al. (2003) and Arkhipova et al. (2009) found a dominant bimodal behaviour for HD\,179821 of $\sim$\,200$^{\rm }$ d and $\sim$\,140$^{\rm }$ d for the time interval 1994--2000. Their ratio is 0.7. The authors suggest that they represent a fundamental tone and the first overtone, respectively, being indicative of a high stellar mass 15--20\,M$_{\rm \odot}$ and T$_{\rm eff}$\,=\,6300\,K according to theoretical models of Zalewski (1986). No reliable result could be obtained for the data between 2000 and 2008 (Arkhipova et al. 2009), while between 2009 and 2017 the quasi-periods increased to more than 250\,d according to Ikonnikova et al. (2018).

\section{2.6. Ratios Ampl\,$V$/Ampl\,$(B-V)$, and the influence of enhanced mass loss episodes.}
References of photometric $\rho$\,Cas data used for this analysis are: the multi-colour photometric observations made by Henry (1995, 1999); Leiker \& Hoff (1987); Leiker et al. (1988, 1989, 1990, 1991); Halbedel (1988, 1991b, 1993a); Zsoldos \& Percy (1991); Percy et al. (2000).

 Figure \,D.1. explains an abnormal amplitude ratio Ampl\,$V$/Ampl\,$(B-V)$  with the aid of three fictitious pulsations in $B$ and $V$.
The first set represents a so-called normal pulsation (see below) as they are the majority, with Ampl\,$V$/Ampl\,$(B-V)$\,=\,1.7\,$\pm$\,0.5. The second set has become smaller according to the observations, mostly in $B$ and less so in $V$ by absorption; see the continuous curves. The corresponding ratio is 3.9. The dashed curves are the intrinsic ones. The third set of amplitudes refers to a `normal' pulsation again without increased absorption, and for which the ratio is `normal': 1.8. To offer a reliable estimate is hardly possible, the inconsistencies in $B$ are 0\fm1 up to perhaps 1$^{\rm m}$, and much less in $V$, depending on the optical thickness of the high-opacity layer.

Caution is called for: we have indications that all pulsations, even `normal' ones, suffer from the same type of absorption, albeit less. Obviously, a high-opacity layer with a variable density seems to be omnipresent. Estimated errors in T(Phot) and radius will be a few 100\,K (too low), and a few 100\,R$_{\rm \odot}$ (too large).

\begin{figure}
 \resizebox{\hsize}{!}{\includegraphics[angle=-90]{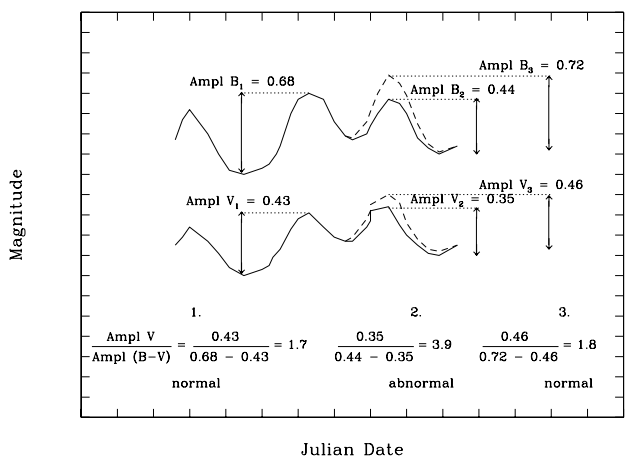}}
   \caption{Example to explain the `cyclic absorption variation' with the aid of a few fictitious pulsations in order to understand the abnormally low amplitudes of the colour curves $(B-V)$; see text.}
    \label{Fig.D.1.}
\end{figure}

\section{2.6. More evidences for a difference between T(Sp) and T(Phot).}
--{\bf $\rho$\,Cas.} The $\rho$\,Cas pulsations 1969--1970 had a mean T(Sp)\,=\,7100\,K (Lobel et al. 1994). The observed mean $(B-V)$\,=\,1.1, after correction for extinction (Table\,1) $(B-V)_{0}$\,=\,0.65, which means a T(Phot)\,=\,5100\,K. This is 2000\,K cooler than T(Sp) mentioned above. Assuming that T(Sp) is correct, the corresponding $(B-V)_{0}$ should be about 0.15. Obviously, the colour $(B-V)$ of $\rho$\,Cas, although in a relatively hot stage, was too red by about 0\fm5 at the time.

--{\bf HD\,179821.} Arkhipova et al. (2009) list in their Table\,5 five spectral temperatures based on CaII and hydrogen P lines, made between 1994 and 2008, using the method of Mantegazza (1991): yielding the average T(Sp)\,=\,6800\,K\,$\pm$\,50\,K.
This agrees with model calculations by Zacs et al. (1996), Reddy \& Hrivnak (1999), and
Kipper (2008), while a 7000\,K is suggested by the far-IR brightness (Nordhaus et al. 2008), and a 7350\,K\,$\pm$\,250\,K by Sahin et al. (2016). However, all these temperatures appeared to be higher than the photometric one for HD\,179821 based on the Straizhys (1982) temperature calibration: 5400\,K--6000\,K. This temperature range for HD\,179821 is of the same order as the one based on the calculated dJN temperatures in the previous sections.

Using observations by Arkhipova et al. (2009), we determined the s-parameter from the $(B-V)_{0}$ according to the dJN calibration, and the photometric T$_{\rm eff}$. We summarise the results: T(Sp)--T(Phot)\,=\,1320\,K and 1210\,K, if the distance is 6\,kpc and 1\,kpc, respectively, with a standard error of the order of 65\,K. Hence, also here spectral temperatures are at least 1000\,K higher than the dJN temperatures.
As the distance difference is extreme, we had to deal with two temperature calibrations, that is, for stars belonging to different luminosity classes: Iab--Ia and IB--II, respectively.

--{\bf HR\,8752.} Nieuwenhuijzen et al. (2012) did not report such differences for HR\,8752, but they do exist and are shown below. This is because these latter authors had the advantage of having of a list of precise MK classifications from which LTE effective temperatures could be derived and subsequently the s-parameter and $(B-V)_{0}$, which could be used to indirectly confirm the existence of that obstinate difference when their temperatures are compared with the ones based on the photometric observations.

Two $BV$ observations from the cool YHG stage of HR\,8752, which were simultaneously observed with T(Sp) determinations:
at JD\,24\,43737 (May 1979), T(Sp)\,=\,5602\,K and T(Phot)\,=\,4880\,K, difference\,=\,722\,K,
and at JD\,24\,45898 (October 1984), T(Sp)\,=\,5343\,K and T(Phot)\,=\,5160\,K, the difference\,= 183\,K. See Tables\,1 and 4 and Fig.\,2  of Nieuwenhuijzen et al. (2012) for references and errors. Photometry by Walker (1983) and Zsoldos \& Olah (1985).

--{\bf HR\,5171A.} AMBER/VLTI observations of HR\,5171A by Wittkowski et al. (2017a), yielded the temperature of T$_{\rm eff}$\,=\,4290\,K\,$\pm$\,760\,K at JD\,245\,6755\,=\,2014. This date point is indicated by a vertical line piece in the lower panel of Fig\,3, and is arbitrarily placed as the $V$ magnitude is unknown; it is situated within the series of $BV$ photometry (Johnson system) and $RI$ photometry (Cousins system) represented by plus symbols in Fig.\,3, of which the data are listed in Table\,I.1. One observation of Table\,I.1. (JD\,24\,56886) was made only 131 d after the temperature determination of Wittkovski et al. It appears that with the dJN method, the temperature of HR\,5171A is T$_{\rm eff}$\,=\,4210\,K, thus well within the error bar of the Wittkovski et al. temperature.
Indeed, inconsistencies in temperature determinations become smaller for lower temperatures.

In this context it is very relevant to mention that Chesneau et al. (2014) suggested the presence of a special veiling envelope rendering a decoupling of photometric and spectroscopic observations. For example, these latter authors showed that  CO molecular lines were affected by this veiling, and could not be used for the spectral classification (their Fig.\,10).
They also concluded that the strong veiling of the K-band spectrum of HR\,5171A was the reason why a spectral classification was impossible. However, here the circumstellar envelope is supposed to be the culprit.

It can be concluded that the results in for example Sect.\,2.8. support the permanent presence of a high-opacity layer causing a variable selective absorption (Appendix Fig.\,D.1.). Its presence is almost undisputable. It is more optically thick for $B$ (a rough estimation: 0\fm1 up to one magnitude) than for $V$ (but the extinction law is uncertain and will be the focus of further research). Naturally the absorption has almost no effect on the spectroscopy, and as a result the differences T(Sp)--T(Phot) increase with temperature. It also appeared that the higher the T(Sp), the higher the gas density of that layer (the higher the selective absorption), meaning that a connection with increased mass loss is almost certain.
See Sect.\,4 for a reference on transparency problems in stellar atmospheres, and for further discussion on the subject.

\section{2.7. Temperature scales.}

A short discussion is needed on a number of temperature calibrations based on broadband photometry.
A new temperature calibration was introduced by de Jager \& Nieuwenhuijzen (1987) and is valid for all main sequence stars up to the hypergiant luminosity class Ia$^{+}$. This calibration is based on the continuous spectral variable `s' derived from $(B-V)_{0}$ for all luminosity classes from V to Ia$^{+}$. This s-parameter determines T$_{\rm eff}$ and the BC. As a result, temperatures for Ia$^{+}$ stars turn out to be cooler, and stellar radii larger than for Iab supergiants, as expected.

An additional novelty of the dJN calibration is, that it allows one to differentiate smoothly between the luminosity classes by means of the variable b-parameter, b\,=\,0.0 for Ia$^{+}$, 0.6 for Ia, and 1.0 for Iab, and so on, to 0.5 for class V.

As the Big Three, namely HR\,8752, $\rho$\,Cas, and HR\,5171A (having similar spectra, Lobel et al. 2015), are without any doubt of class Ia$^{+}$ (with b\,=\,0), their bolometric magnitudes should vary during the pulsations, and considering their spectral types, be roughly between -8.7 and -9.6. If our calculations result in a slightly low luminosity, we neglect them and still consider the star as a Ia$^{+}$ YHG with b\,=\,0. Our consideration is that also uncertainties in distances and reddening are not negligible at all.

We found and described in Sects.\,2.4.--2.6. and Appendix E:\,2.6. that photometric temperatures for YHGs are always lower than spectroscopic temperatures, and offer an explanation.

We also tried to derive temperatures from theoretical spectral energy distributions (T(SED)), by comparing the observed SEDs based on the photometric fluxes between two channels (e.g. between $B$ and $V$, or $V$ and $R$, etc.), corrected for interstellar extinction, with the corresponding theoretical fluxes of atmospheric models (log\,g\,=\,0--0.5, v$_{\rm turb}$\,=\,2\,kms$^{-1}$, solar abundance) by Castelli \& Kurucz (2003) and Castelli (2014).

Our test case was $\rho$\,Cas using $BV$ and $VRI$ photometry by Henry (1995, 1999; see Sect.\,2.2.) coinciding 6 T(Sp) values from Klochkova et al. (2014) and 4 T(Sp) from Lobel et al. (2003). The extinction and reddening used are listed in Table\,1. It turned out that the differences $\triangle$\,T\,=\,T(Sp)--T(SED) for each set of $BV$, $VI,$ and $VR$ are linearly related with the observed brightness $V$ (the three running parallel), and are thus roughly related with the temperature: the brighter the star, the larger the difference between T(Sp) and T(SED), the latter being lower. For $BV$ the difference grows from $\sim$\,200\,K near $V$\,=\,4\fm72 (the coinciding T(Sp) determinations only cover a small brightness range), to $\sim$\,800\,K near $V$\,=\,4\fm5. For $VI$ these numbers are from $\sim$\,0\,K near $V$\,=\,5\fm1 to $\sim$\,2500\,K near $V$\,=\,4\fm2. For $VR$ these numbers are from $\sim$\,600\,K near $V$\,=\,5\fm1 to $\sim$\,2900\,K near $V$\,=\,4\fm2.

These inconsistencies, increasing with temperature, point to an increasing flux excess of the observations with respect to the models in the sense that $V$ shows slightly more flux than $B$, $R$ more than $V$, and $R$ much more than $I$. These relative excesses transformed into magnitudes range from $\sim$\,0\fm1 to 0\fm7.
We cannot offer a plausible explanation. It may be that the models for such extreme stars are not yet perfect. Therefore, we disregarded the T(SED) results.

\section{2.8. Monitoring of V$_{\rm rad}$ is indispensable.}

Individual pulsations are unique, each (at least most of them) can be considered more or lesl a global feature: the entire surface moves up and down, albeit not necessarily spherically symmetric as we are probably dealing with non-radial pulsations. This is supported by the radial velocity monitoring mentioned above of $\rho$\,Cas and HR\,5171A, revealing an oscillating trend in pace with the light variations without significant irregularities (as far as the accuracy of the current radial velocity curves allow such a statement), and showing a stable phase lag of about 0.4 (Lobel et al. 2003, 2015).
Minimum radius happens without any doubt always at maximum brightness and vice versa (as far as the available measurements allow such an assertion). This in contrast with the calculated properties for many YHG pulsations and based on multi-colour photometry, which is deteriorated by atmospheric absorption.

An explanation for the higher reliability of radial velocity measurements is as follows. They are mainly governed by the central part of the pulsating sphere, and are much less affected by the outer ring suffering from projection effects and limb darkening, or by a variable non-circular and irregularly shaped circumference (as seen from the Earth), see recent images of HR\,5171A by Wittkovsky et al. (2017c) discussed in Sect.\,3.4.3. of the present paper. Additionally, one expects little disturbing effects on the measurements by dynamically active surface features or variable absorbing effects by the atmosphere. On the contrary, multi-colour photometry is highly sensitive to most of these factors.

Therefore, we would like to stress that long-lasting monitoring of V$_{\rm rad}$ curves, especially with a small time resolution, is indispensable for deeper studies, and was a regrettable deficiency of the present analysis. It should be noted that radial velocity curves also show a long-term variation more or less in accordance with the LTVs (Figs.\,2 and 3 in Lobel et al. 2003).

\section{2.9. Distance of HR\,5171A}
We attempted to calculate the stellar properties of HR\,5171A in a similar way as we did for two pulsations of $\rho$\,Cas (Tables\,2 and 3), this time in order to compare V$_{\rm rad, puls}$ with $V_{\rm rad}$, the spectroscopic velocity. The goal was to find the distance.

Our intention was to compare both radial velocity amplitudes, the one based on the geometrical velocities of the stellar surface V$_{\rm rad, puls}$, with aid of calculations similar to those for $\rho$\,Cas (Table\,3), and the other based on the atmospheric spectral lines (10\,kms$^{-1}$ mentioned in Sect.\,2.9.) and multiplied by $\sim$\,1.4 (to correct for projection and limb darkening), hence about 14\,kms$^{-1}$. The purpose was to determine the true distance of HR\,5171A. After all, radius, radius variation, and radial pulsation velocity are distance dependent.
 When the latter (in kms$^{-1}$) is plotted versus a number of assumed distances used for the calculations, a linear relation is obtained. The right distance should then be where V$_{\rm rad, puls}$\,=\,V$_{\rm rad}$ (corrected)\,=\,14\,kms$^{-1}$.
For the calculations we used distances of 3.6\,kpc down to 1\,kpc. The attempt failed, most likely due to inconsistent $V$ and $(B-V)$ parameters (just like in the case of the pulsations of $\rho$\,Cas in Tables\,2 and 3, and of HR\,8752 and HD\,179821 discussed in Sect.\,2.4.--2.6.).

Yet the ratio Ampl\,$V$/Ampl\,$(B-V)$ of the ascending branch of HR\,5171A above was not exceptionally high: 1.8, the one for the descent was high: 2.9. However, due to scarcity of data points, these numbers are not reliable. In any case, as a result of the photometric inconsistencies, calculated radial pulsation velocities appeared to be too small, or ran opposite to the observed V$_{\rm rad}$ values.

Just to get some idea, we list below only the calculated temperature and radius ranges for 1.5\,kpc (temperature similar for 3.6\,kpc). We estimate that the values for T(Phot) are too faint by a few hundred degrees, and the radii are too small by about 100\,R$_{\sun}$ .

Summary for 1.5\,kpc:
dJN: T ranges 4020--4200\,K, and R ranges 1060--1160\,R$_{\rm \odot.}$

The best fit to the optical spectrum of 1992 (in the minimum after maximum no.\,17 (in Paper\,I Fig.1) by Lobel et al. (2015) is for 5000\,K and log\,g\,=\,0. As usual, the spectroscopic temperature is higher than the photometric one above.

This type of distance determination will only be successful when the photometric parameters of a pulsation are consistent.

\section{3.1. Long-term variations}
Long-term variations were noted and discussed by many observers: for HR\,5171A, for example, Harvey 1972, Paper\,I, Chesneau et al. (2014); for $\rho$\,Cas: for example, Arellano Ferro (1985), Percy \& Keith (1985), Leiker et al. (1988, 1989, 1991), Halbedel (1988b, 1991b, 1993a, b) and Zsoldos \& Percy (1991); for HD\,179821: Arkhipova et al. (2001, their Fig\,11) and Le Coroller et al. (2003).

  \begin{table*}
  \caption{The photoelectric observations of HR\,5171A in $BV$ (Johnson) and $RI$ (Cousins) made from 2013 to 2018 by GDS. Six of these observations ($V$ and $(B-V)$) are plotted in Figs.\,13 and 14, and only the $V$ magnitudes in Fig.\,3.}
   \label{table:I.1.}
   \centering
   \begin{tabular}{c|c c c c c} \\
   \hline\hline
   JD-- & & & &   \\
   245\,0000 & $V$ & $(B-V)$ & $(V-R)_{\rm c}$ & $(V-I)_{\rm c}$ & $(R-I)_{\rm c}$   \\
   \hline
    6324.03 & 6.53 & 2.54     \\
    6330.01 & 6.59 & 2.44     \\
    6331.06 & 6.58 & 2.46      \\
    6337.09 & 6.59 & 2.51      \\
    6886.50 & 6.60 & 2.46 & 1.38 & 2.73 & 1.35  \\
    7399.18 & 7.03 & 2.47 & 1.54 & 2.85 & 1.31       \\
    7426.07 & 6.84 & 2.41 & 1.49 & 2.62 & 1.13       \\
    7636.86 & 6.93 & 2.35 & 1.49 & 2.60 & 1.11       \\
    8132.95 & 6.81 & 2.49 & 1.46 & 2.74 & 1.28       \\
    8136.19 & 6.79 & 2.49 & 1.46 & 2.72 & 1.25       \\
    \hline
   \end{tabular}
    \end{table*}

\section{3.4.2. Long-term variation of HR\,5171A in the $VBLUW$ system.}
 Figure\,J.1. shows the detailed smoothed LTVs for the four colour indices in (log intensity scale) of the Walraven $VBLUW$ system (described by Lub \& Pel 1977; de Ruiter \& Lub 1986) for the decade 1980-1991, and derived from Fig.\,2 of Paper\,I. They all show undulating declining reddening trends in $V-B$ (equivalent to the $(B-V)$ of the $UBV$ system), $B-L$ and $B-U$ (is roughly similar to the $(B-U)$ of the $UBV$ system, but due to the large differences in band widths it is impossible to derive transformation formulae). The average declines in especially $V-B$ and $B-U$, confirm the reddening trend of $(B-V)$ in Figs.\,13 and 14. The small bar on the right in the top panel indicates the local brightness minimum in the $VBLUW$ time series.

 Most interesting is the $U-W$ curve being dominated by two cycles of about 2000d mainly due to a $W$ ($\lambda$\,3235\,\AA) brightness variation. The second cycle with a prominent maximum around (1989--1990), amplitude $\sim$\,0.12 log scale, or 0\fm3, points to an increase of the Balmer continuum radiation relative to the observations made in 1971, 1973, 1980 and 1981 (see the two-colour diagrams in Fig.\,5 in Paper\,I).
 It is also striking that some of the highest amplitudes of the pulsations in the near-UV pass bands $L$, $U,$ and $W$ (Fig.\,2 in Paper\,I) happen during these two $U-W$ maxima in Fig.\,J.1., roughly around JD\,24\,45200 and JD\,24\,47600.
 We presume that this undulating Balmer continuum radiation is somehow related to the circumstellar matter (Sect.\,2.9.).

 \begin{figure}
   \resizebox{\hsize}{!}{\includegraphics[angle=-90]{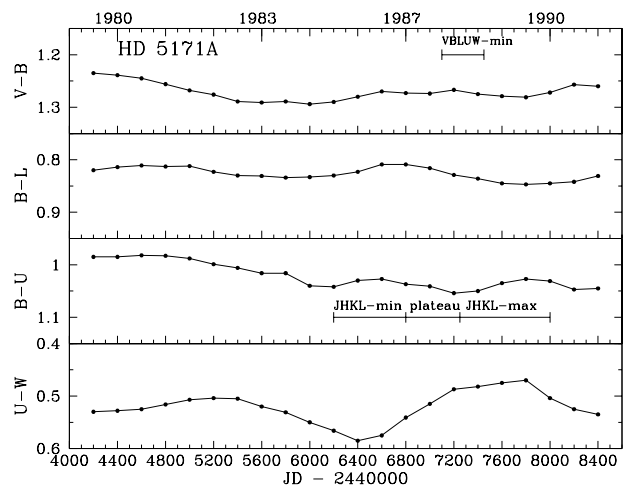}}
   \caption{Long-term variations of HR\,5171A in the four colour indices (in logarithmic scale) of the medium broadband $VBLUW$ system 1980--1991, showing a slightly undulating
  reddening trend for the $V-B$ (equivalent to the $(B-V)$ of the $UBV$ system), $B-L$ and $B-U$ colour indices. Blue is up. The $U-W$ curve behaves differently, showing two
 oscillations due to Balmer continuum brightness changes with a timescale of about 2000$^{\rm }$d. The last maximum ($\sim$\,JD\,24\,47600) has an amplitude of about 0\fm3. This maximum and the preceding minimum might be related to a local maximumm or plateau and a minimum, respectively, in the $K$ light curve (thus, also likely present in the other NIR bands). Bars labeled with `$JHKL$-etc' in the $(B-U)$ panel will be discussed in the next paper.}
    \label{Fig.J.1.}
    \end{figure}

\begin{table*}
  \caption{The light and colour index ranges of the LTVs, shown in Figs.\,6--14.}
   \label{table:J.1.}
   \centering
    \begin{tabular}{c|c c c c c c c c c c}
   \hline\hline
   star & $V$ & $B$ & $U$ & $R$ & $I$ & $B-V$ & $U-B$ & $R-I$ & $V-I$ & $R-I$  \\
  dates &     & ($B_{\rm pg}$)   \\
  \hline
  HR\,8752                                \\
  1940--1994 & 0.6 & 0.8 & 1.2 & & & 0.9 & $>$0.7           \\
  \hline
   HR\,5171A                                            \\
   1900--1950 & & (1.3)                                  \\
   1953--2018 & 0.9 & 1.2 & & & & 0.7                        \\
   \hline
   $\rho$\,Cas                                            \\
  1963--2018 & 0.5 & 0.5 & 0.8 & 0.3 & 0.2 & 0.4 & 0.6 & 0.2 & 0.4 & 0.2   \\
  \hline
  HD\,179821                                               \\
  1899--1989 &  & (0.5)                                    \\
  1990--2009 & 0.15 & 0.12 & 0.4 & & & 0.2 & 0.3                 \\
  \hline
  \end{tabular}
  \end{table*}

\section{3.3. Trends of brightness, quasi-periods, and amplitudes of HR\,8752 during the BL evolution 1993-2019.}
The trends of amplitudes and quasi-periods of HR\,8752 were described in Sect.\,3.1.
We explored the AAVSO website for further information, not only on the trends of amplitudes and quasi-periods, but also on the brightness. The scatter of the AAVSO mean light curves sometimes hampered the determination of a significant mean quasi-period (also due to the low intrinsic amplitudes of about 0\fm1). However, AAVSO photometric $V$ (Johnson) observations were made as well and were a great help. After 1993 ($V$\,$\sim$\,5.05), all $UBV$ monitoring stopped for some unknown reason.

The photometric $V$ magnitudes were usually about 0\fm2--0\fm4 brighter than the visual estimates (as expected). The mean brightness $V$ and estimated amplitudes were as follows, interval 1993--1995: 5.15 and 0\fm1--0\fm2; 1995--1998: 5.2 and 0\fm1; 1998--2001: 5.25 and 0\fm05; 2002--2005: 5.3; 2005--2008: 5.3; 2008--2011: 5.3 and 0\fm1; 2011--2014: 5.25; 2012--2018: 5.3 and 0\fm1.
The conclusion is that the brightness $V$ of HR\,8752 stayed almost constant from the early 1990s until 2019 (the finishing date of this paper). The light amplitudes (if real) were of the order of 0\fm1.

\section{5. Message to observers}
Young professionals should start a new era of worldwide coordinated observing campaigns for YHGs (e.g. with the photometric and spectroscopic opportunities offered by the Las Cumbres Observatory, a global telescope network of robotic telescopes (headquarters in Goleta, CA, USA).

 While the AAVSO observers continued observing, most of the professionals stopped (or had to stop) observing HR\,5171A and HR\,8752 all of a sudden in the 1990s. Considering the unprecedented number of exciting and spectacular activities of YHGs, and not to forget all other types of variables, with respect to their physics, instability, and evolution, this halt 30 years ago was deplorable, supporting Sterken's (2002: Sect.\,3.1.) concern on the same matter.

Hopefully, our results herald coordinated campaigns for long-lasting spectroscopic monitoring of YHGs for obtaining detailed radial velocities, spectral characteristics, T(Sp), and so on. Naturally, such campaigns would go hand in hand with multi-colour photometry, preferably with a time resolution of a week or less. Only then can reliable information on for example the nature of the high-opacity layer and the absorption law be obtained, as well as information on the detailed development of eruptive episodes and on the subtle changes of pulsating and stellar properties along the BL and RL evolutions.

\section{Electronic tables for the photometric observations of $\rho$\,Cas.}

\begin{table*}
 \caption{Stub-table for the full T2 APT $VRI$ photometric dataset of $\rho$\,Cas, 1986--2001, in electronic form available at the CDS (Sect.\,2.2). See Henry (1995, 1999) for a description of the T2 0.25\,m Automatic Photoelectric Telescope (APT) operations and data reduction. Magnitudes (mag) of variable (Var) and check (Chk) star relative to the comparison star HD\,223173. A `99.999' signifies that the differential magnitude was discarded because its internal standard deviation exceeded 0\fm02.}
  \label{table: M.1.}
   \centering
   \begin{tabular}{c c c c c c c}
  \hline\hline
  Date &  Var $V$  & Var $R$ & Var $I$ &  Chk $V$ & Chk $R$ & Chk $I$  \\
  (HJD - 2\,400\,000) & (mag) & (mag) & (mag) & (mag) & (mag) & (mag)    \\
  \hline
  46679.8629 & -0.840 & -0.731 & -0.605 & 99.999 & 99.999 & 99.999 \\
  46679.9081 & -0.838 & -0.720 & 99.999 & 0.738  & 1.685  & 2.323  \\
  46680.7515 & -0.839 & -0.741 & -0.605 & 0.728  & 1.666  & 2.318  \\
  46680.8244 & -0.850 & -0.722 & -0.607 & 0.737  & 1.690  & 2.317  \\
  46680.8862 & -0.845 & -0.749 & -0.626 & 0.746  & 1.660  & 2.339  \\
  \hline
  \end{tabular}
  \end{table*}

\begin{table*}
 \caption{The same as Table\,M.1., but now for the full T3 APT $BV$ photometric dataset of $\rho$\,Cas, 2003--2018. A `99.999' signifies that the differential magnitude was discarded because its internal standard deviation exceeded 0\fm01.}
 \label{table: M.2.}
  \centering
   \begin{tabular}{c c c c c }
  \hline\hline
  Date  & Var $B$ & Var $V$ & Chk $B$ & Chk $V$ \\
  (HJD - 2\,400\,000 & (mag) & (mag) & (mag) & (mag) \\
  \hline
  52800.9656  & -1.261  & -0.921 & -0.402 & 0.840 \\
  52807.9611  & -1.272  & -0.925 & -0.401 & 0.853 \\
  52828.9681  & -1.282  & -0.939 & -0.402 & 0.848 \\
  52894.7499  & -1.273  & -0.950 & -0.405 & 0.830 \\
  52895.6689  & -1.286  & -0.957 & -0.409 & 99.999 \\
   \hline
   \end{tabular}
   \end{table*}

   \end{appendix}

   \end{document}